\documentclass{elsart_sf}
\usepackage{epsfig,subfigure,multirow}
\begin{document}
\begin{frontmatter}
\title{Transmission of light in deep sea water at the site of the {\sc Antares} 
neutrino telescope}

\renewcommand{\thefootnote}{\fnsymbol{footnote}}
\begin{center}          
{\bf The ANTARES Collaboration}\\
\vspace{1mm}
\author[ific]{J.A.~Aguilar},
\author[grphe]{A.~Albert},
\author[lam]{P.~Amram},
\author[dip-genova]{M.~Anghinolfi},
\author[erlan]{G.~Anton},
\author[dapnia]{S.~Anvar},
\author[dapnia]{F.E.~Ardellier-Desages},
\author[cppm]{E.~Aslanides},
\author[cppm]{J-J.~Aubert},
\author[dapnia]{R.~Azoulay}
% ...
\end{center}
\address[ific]{IFIC -- Instituto de F\'{\i}sica Corpuscular, Edificios Investigaci\'on de Paterna, CSIC -- Universitat de Val\`encia, Apdo. de Correos 22085, 46071 Valencia, Spain}
\address[grphe]{GRPHE -- Groupe de Recherches en Physique des Hautes Energies, Universit\'e de Haute Alsace, 61 Rue Albert Camus, 68093 Mulhouse Cedex, France}
\address[lam]{LAM -- Laboratoire d'Astrophysique de Marseille, CNRS/INSU - Universit\'e de Provence Aix-Marseille I, Traverse du Siphon -- Les Trois Lucs, BP 8, 13012 Marseille Cedex 12, France}
\address[dip-genova]{Dipartimento di Fisica dell'Universit\`a e Sezione INFN, Via Dodecaneso 33, 16146 Genova, Italy}
\address[erlan]{University of Erlangen,Friedrich-Alexander Universit\"at Erlangen-N\"urnberg, Physikalisches Institut,Erwin-Rommel-Str. 1, 91058 Erlangen, Germany}
\address[dapnia]{DSM/DAPNIA -- Direction des Sciences de la  Mati\`ere, D\'epartement d'Astrophysique de Physique des Particules de  Physique Nucl\'eaire et de l'Instrumentation Associ\'ee, CEA/Saclay, 91191 Gif-sur-Yvette Cedex, France}
\address[cppm]{CPPM -- Centre de Physique des Particules de Marseille, CNRS/IN2P3 Universit\'e de la M\'editerran\'ee Aix-Marseille II, 163 Avenue de Luminy, Case 907, 13288 Marseille Cedex 9, France} 
% ...
\address[dip-bologna]{Dipartimento di Fisica dell'Universit\`a e Sezione INFN, Viale Berti Pichat 6/2, 40127 Bologna, Italy}
\address[dip-bari]{Dipartimento Interateneo di Fisica e Sezione INFN, Via E. Orabona 4, 70126 Bari, Italy} 
% ...
\address[dip-catania]{Dipartimento di Fisica ed Astronomia dell'Universit\`a e Sezione INFN, Viale Andrea Doria 6, 95125 Catania, Italy}
% ...

\begin{abstract}
The ANTARES neutrino telescope is a large photomultiplier array
designed to detect neutrino-induced upward-going muons by their
Cherenkov radiation. Understanding the absorption and scattering of
light in the deep Mediterranean is fundamental to optimising the
design and performance of the detector. This paper presents
measurements of blue and UV light transmission at the ANTARES site
taken between 1997 and 2000. The derived values for the scattering
length and the angular distribution of particulate scattering were
found to be highly correlated, and results are therefore presented 
in terms of an absorption length $\lambda_{\mathrm{abs}}$ and an
effective scattering length $\lambda_{\mathrm{sct}}^{\mathrm{eff}}$.
The values for blue (UV) light are found to be
$\lambda_{\mathrm{abs}} \simeq 60(26)$ m, 
$\lambda_{\mathrm{sct}}^{\mathrm{eff}}     
\simeq 265(122)$ m, with significant ($\sim$15\%) time
variability.  Finally, the results of ANTARES simulations showing
the effect of these water properties on the anticipated
performance of the detector are presented.

\end{abstract}

\vspace{2cm}
\end{frontmatter}
\normalsize
\newpage


\section{Introduction}

\begin{figure}[h] \begin{center}
  \epsfig{file=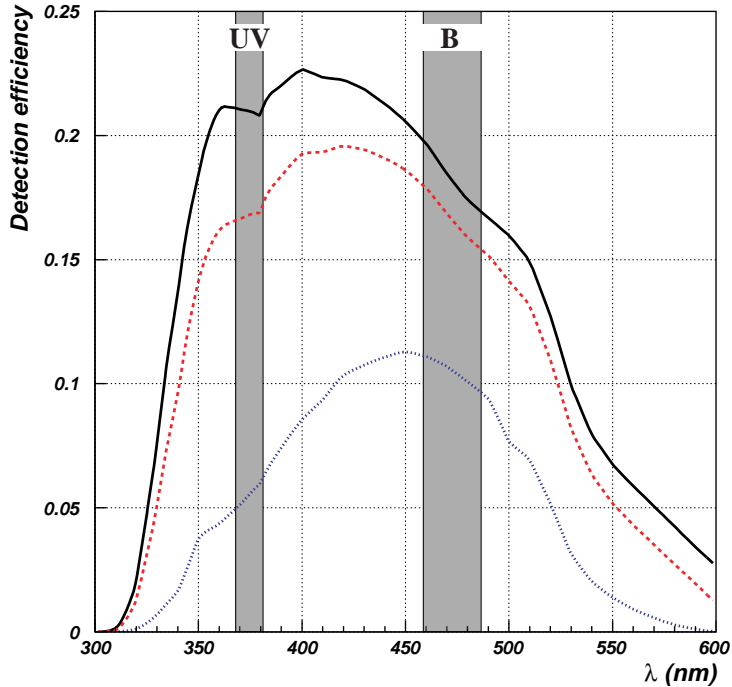,width=.7\textwidth} \caption{Detection
  efficiency as a function of wavelength.  The solid curve shows the PMT quantum efficiency and the absorption by the glass
  of the PMT, the optical gel and the protective glass sphere housing
  the PMT~\cite{Pro}. The dashed and dotted curves are calculated with a path length in water
  of 5~m and 30~m respectively, assuming a characteristic wavelength dependence 
  of the water
  absorption length as given in~\cite{Price}. The
  effect of scattering is not included. The bands labelled ``B'' and
  ``UV'' indicate the wavelengths at which measurements were
  undertaken at the {\sc Antares} site.}  \label{fig:att} \end{center}
\end{figure}

The {\sc Antares}\footnote{http://antares.in2p3.fr} undersea neutrino telescope~\cite{WEB} will use
an array of photomultiplier tubes  (PMT) in the deep Mediterranean Sea to detect
the Cherenkov light emitted by muons resulting from the interaction
of high energy neutrinos with matter.  The muon
track is reconstructed from the arrival time of detected photons. The performance of
the detector is therefore critically dependent on the optical
properties of sea water, in particular on the velocity of light and on
the absorption and scattering cross-sections.  All these parameters
vary with the photon wavelength. The relevant spectrum spans from ultraviolet
to green (see figure~\ref{fig:att});  the Cherenkov light spectrum varies like $1/\lambda^2$, the photomultiplier tube
quantum efficiency becomes too low to probe wavelengths longer than
$600$~nm, while the glass pressure sphere that surrounds the
phototube absorbs the light at wavelengths shorter than $320$~nm.
Seasonal variations in sedimentation in the sea water~\cite{fouling} might induce variations in the optical parameters.

Several measurements of sea water attenuation have been performed in the past. The Dumand collaboration reported an attenuation length varying with the light wavelength and reaching a maximum value of 60~m, with 50\% accuracy, for a wavelength of 500~nm~\cite{Dumand}. The Nestor collaboration measured a similar behaviour. The maximum attenuation length was found at 490~nm and was estimated to be $55\pm10$~m~\cite{Nestor}. The Ba\"ikal experiment found a maximum absorption length of about 20~m for a wavelength of 490~nm~\cite{Baikal}. Measurements performed in pure water show that the maximum of the attenuation length is at lower values of the wavelength in this medium ($\sim 400$  nm)~\cite{purewateratt1,purewateratt2}. The maximum value was measured around 90~m with 40\% accuracy. More recently, measurements of the absorption spectrum in pure water have been reported, with maximum absorption lengths of 160 $\pm$ 15 m at 420 nm~\cite{purewaterabso1} and of 225 $\pm$ 30 m at 417 nm~\cite{purewaterabso2}.

In order to reach an optimal knowledge of the light propagation
properties at the detector site all relevant parameters 
concerning photon absorption and scattering should be measured.
These parameters, described in section~\ref{sec:sim}, will be directly measured and continuously monitored by the
{\sc Antares} experiment using an instrumentation
line.  Until now, the adopted approach has been to measure these parameters with {\em in situ} autonomous devices, and these measurements are the subject of this paper. 
We present time-of-flight 
distributions of photons emitted from a pulsed isotropic light source and 
detected by a PMT at different distances from the source and for two 
wavelengths (blue and UV, as indicated in figure~\ref{fig:att}).  Knowledge of the time-of-flight distribution is essential in order to reconstruct 
muon tracks. While this approach is not sufficient to fully determine the 
differential cross section of the photon scattering process, the absorption 
length can be measured unambiguously. A 
parameterisation which reproduces the main 
features of the scattering process can be obtained, sufficient for the needs of 
the tracking  algorithms and of the detector simulation. 


\section{Experimental setup and measurement procedure}

The site chosen for the deployment of the detector is southeast of
Toulon ($42^\circ 50'$N $6^\circ 10'$E), 40~km from
shore at a depth of 2475~m.  During several sea campaigns from 1997 to
2000 we have improved the experimental setup devoted to the study of
the light transmission properties, and refined the analysis of the
data.  We focus, in this section, on the final experimental
configuration.


\subsection{The mooring line}\label{sec:line}
\begin{figure}[h] \begin{center} 
  \epsfig{file=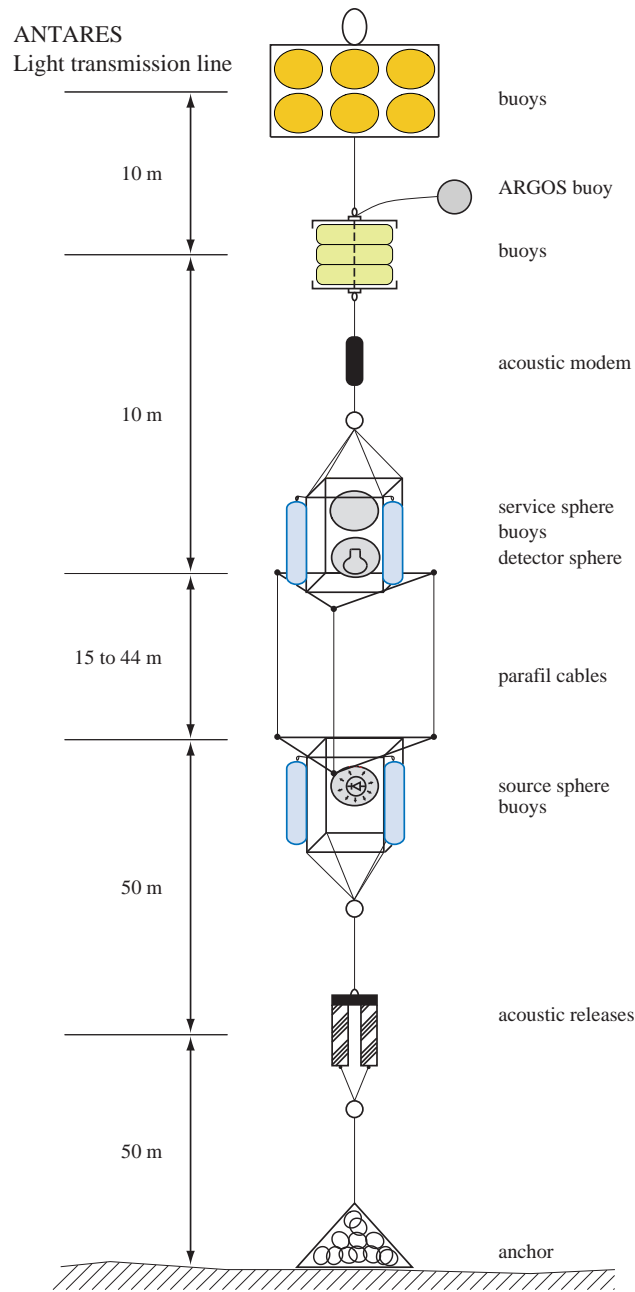,height=.75\textheight} \caption{Sketch of
  the mooring line used for the measurements of the water transmission
  properties. The figure is not to scale.} \label{fig:line}
  \end{center}
\end{figure}

The measuring system is mounted on an autonomous mooring line anchored
by a sinker.  The line remains vertical through the flotation provided
by syntactic buoys.  After deployment, an acoustic modem\footnote{ATM
845/851 from Datasonic (now Benthos), www.benthos.com} is used to
control the measurement from the surface ship which stays in the
vicinity of the zone where the line was sunk. A sketch of the mooring line
including information on approximate heights from the sea floor is
displayed in figure~\ref{fig:line}.

The measuring system consists of 17$''$ pressure resistant glass
spheres mounted on two triangular aluminum frames.  A set of three
mechanical cables attached to the vertices of the two frames defines
their separation distance.  The bottom frame supports a light source
sphere which contains a set of LEDs with their pulsers.  The top frame
supports a detector sphere facing the light source sphere, and a
service sphere (cf. figure~\ref{fig:elect}). 
The detector sphere houses a photomultiplier tube,
the DC-DC converter which supplies the high voltage and a pulse height
discriminator providing a timing signal for each detected photon.  The
service sphere contains a TDC, a microprocessor which controls the
measurement and records the data, and a set of lithium batteries which
power the system.  An acoustic modem remote unit is located on top of
the measuring system.  
 

\subsection{The light source sphere}\label{sec:source}

\begin{figure}[hhh] \begin{center}
  \mbox{\subfigure[UV LED]
  {\epsfig{file=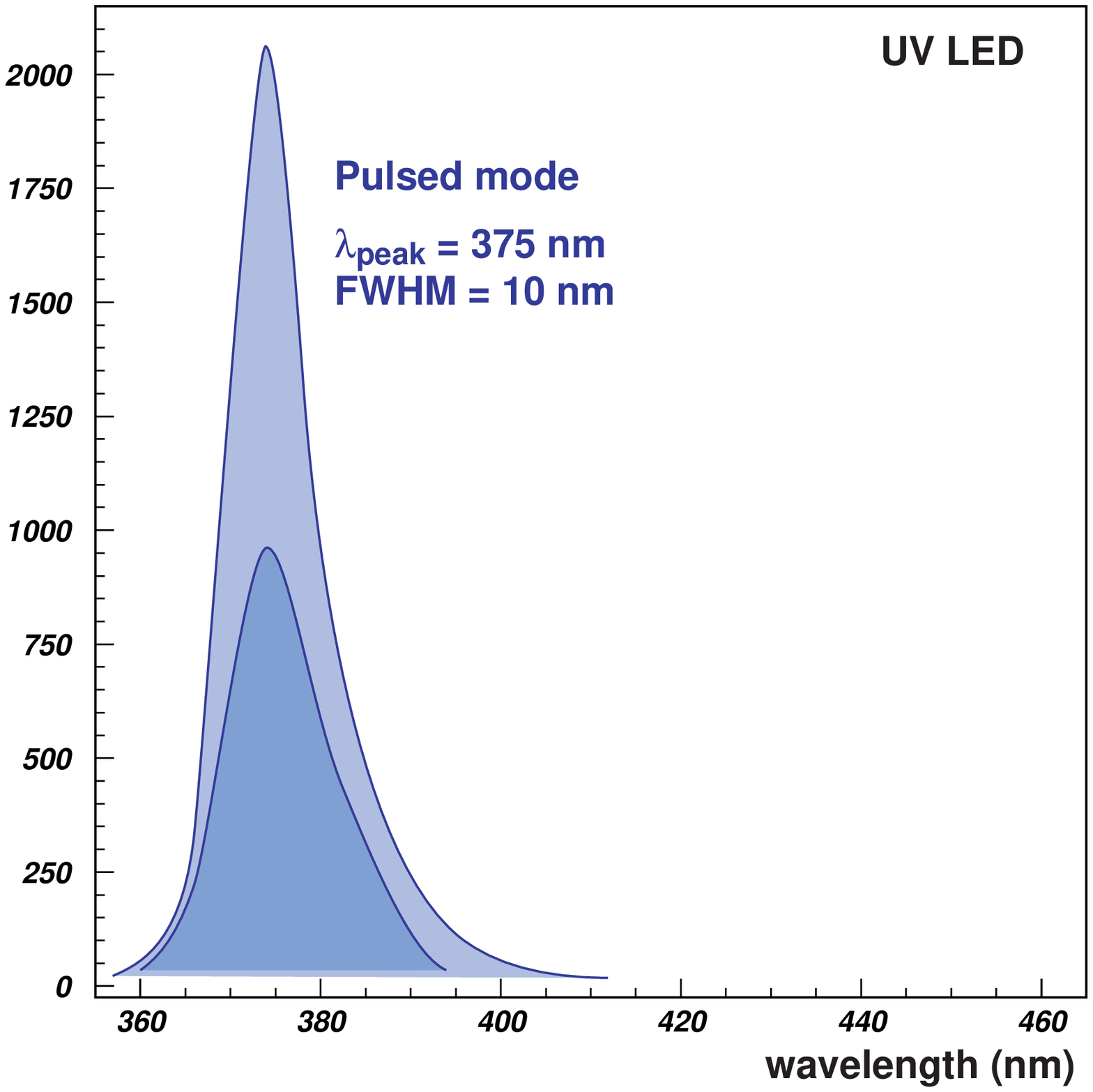,width=.49\textwidth}
  \label{fig:sourceUV}} \quad \subfigure[Blue LED]
  {\epsfig{file=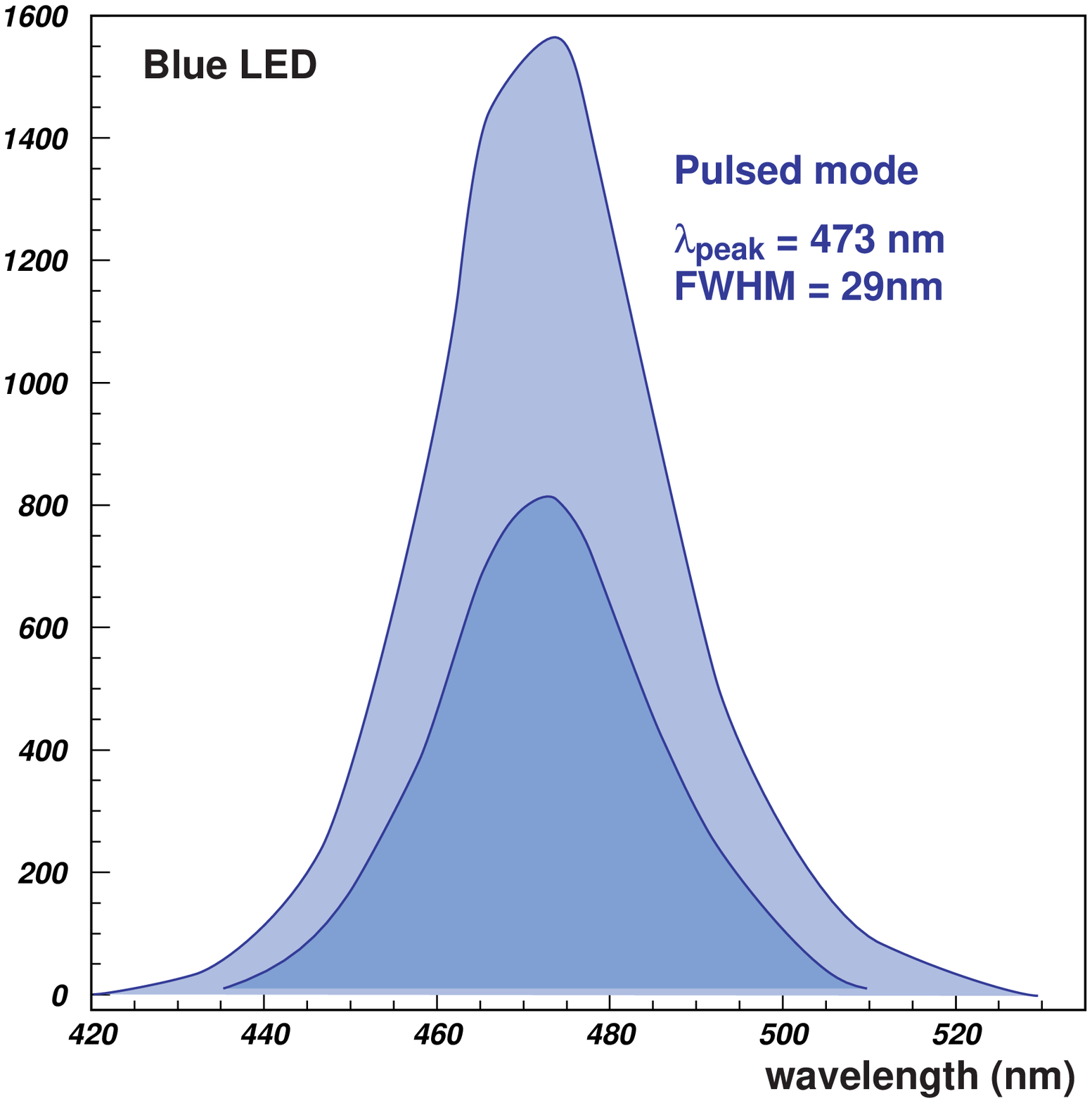,width=.49\textwidth} \label{fig:sourceB}}}
  \caption{Spectral emission of the blue and UV LEDs (smoothed curves, intensity in arbitrary units).  The light and dark areas correspond to the two extreme values of the range of source intensities used for the measurements, with no significant difference in the spectra.}  \label{fig:sources}  \end{center}
\end{figure}
\begin{figure}[hhh] \begin{center}
  \epsfig{file=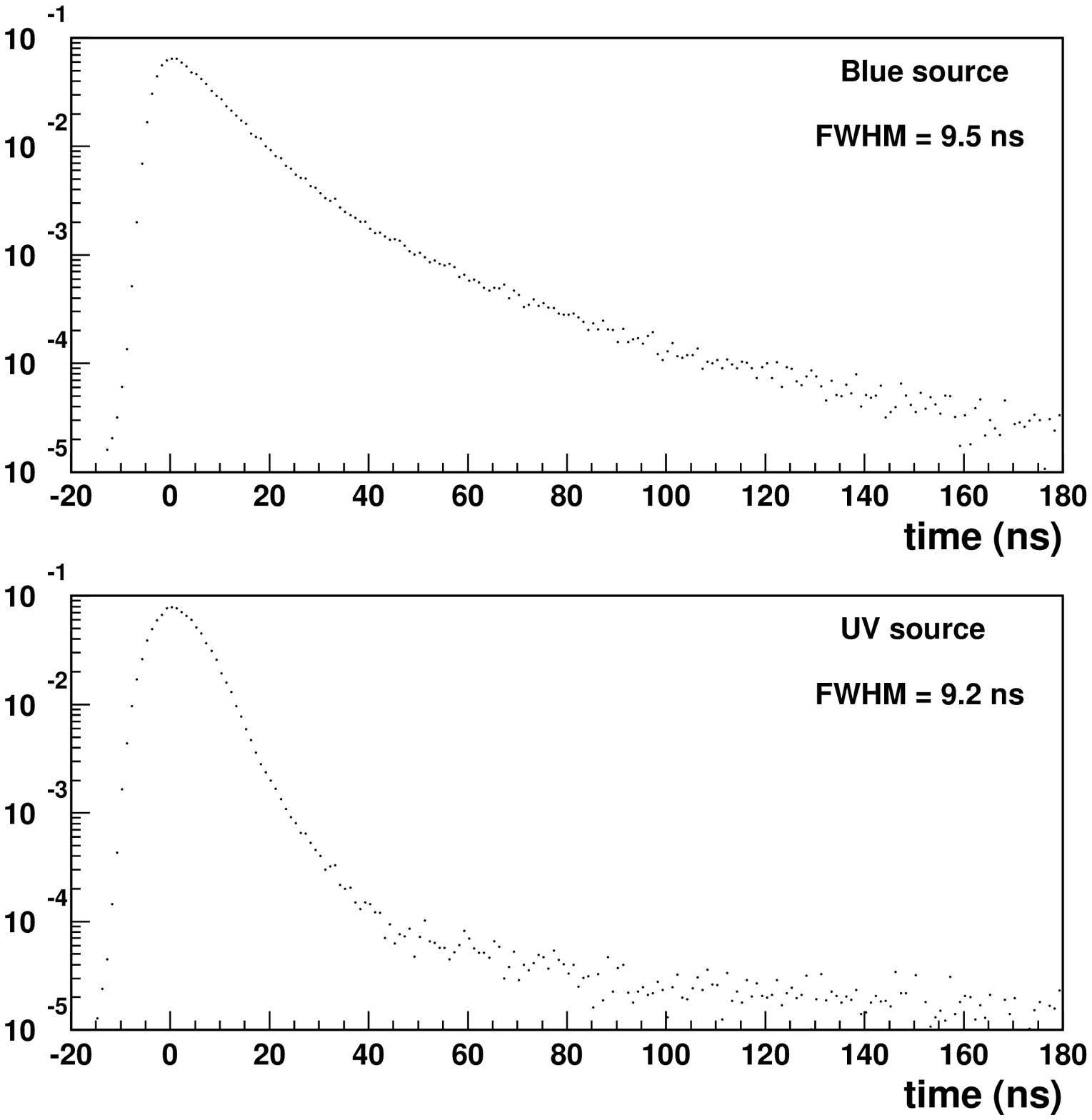, width=.7\textwidth} \caption{Normalised distributions of photon arrival times 
  measured in air with high statistics for the complete source sphere
  and with the full setup. Collimators are located between the source
  and the detector spheres.} \label{fig:air}  \end{center}
\end{figure}

In order to obtain an isotropic light source for two wavelengths, 6 
pairs of LEDs were mounted on the centres of the faces of a cubic 
frame 3~cm on a side which also supports the LED pulser boards.  Each 
pair, which includes a blue and a UV LED, is covered with a single 
1~cm diameter diffusing cap consisting of glass micro-spheres embedded 
in epoxy.  The cube is installed at the centre of a 17$''$ glass 
sphere whose external surface has been sand blasted to provide extra 
diffusion and to remove surface ripples or roughness which can destroy 
the homogeneity of the emitted light flux.  The blue or UV emission 
colour is chosen for each new acquisition by the operator; all 6 LEDs of 
the selected colour are then flashed simultaneously.

The spectrum of the light emitted by the LEDs was measured using a
spectro-photometre (see figure~\ref{fig:sources}).  The peak wavelength
and the spectrum FWHM in pulsed mode operation are (375~nm, 10~nm) and
(473~nm, 29~nm) respectively for the UV and blue LED. The time distribution 
of photons emitted by the light source
sphere was measured in a dark room for both colours using the complete
system (see figure~\ref{fig:air}) and has a FWHM of about 9~ns, with a
tail towards longer times.  The isotropy of the source
was checked to be within $\pm$12\% by measuring the light flux
for different orientations of the source sphere with respect to the
detector sphere.


\subsection{The detector sphere}

A small 1$''$ diameter photomultiplier tube\footnote{Photomultiplier
tube 9125 SA from EMI, now ETL,\\ www.electron-tubes.co.uk/splash.html}
glued on the internal surface of a 17$''$ sphere detects photons
emitted by the light source.  The size of the PMT was chosen in
order to limit the counting rate from deep sea optical background. 
The PMT was then selected for its speed and low transit time spread.
Except for the PMT window, the internal surface of the detector sphere
is blackened in order to absorb photons outside the PMT detection
solid angle.  


\subsection{The service sphere}\label{sec:service}

\begin{figure}[h] \begin{center} 
  \epsfig{file=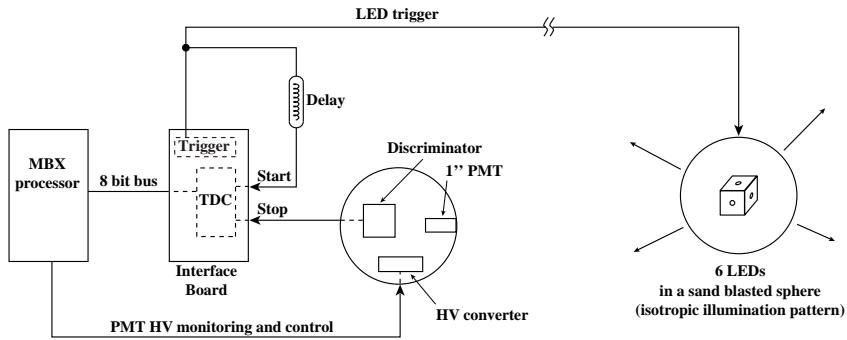,width=.8\textwidth} \caption{Sketch of the
  acquisition system.} \label{fig:elect} \end{center}
\end{figure}

The service sphere provides a 6 kHz trigger signal which is fed to the
LED pulsers and through a delay to a TDC (cf. figure~\ref{fig:elect}).
The TDC is started by the delayed trigger signal and stopped by the first
PMT signal above the discriminator threshold. The clock of the TDC is
defined by a 40~MHz quartz oscillator with each 25 ns period
subdivided into 32 approximately equal channels giving an average 
$\delta t=0.78$~ns time bin. The TDC range
can be adjusted by defining its active window; during most of the
measurements it was set to 1100 channels in order to accommodate the
time distributions for all of the source-detector distances
investigated.

\begin{figure}[h] \centering
  \epsfig{file=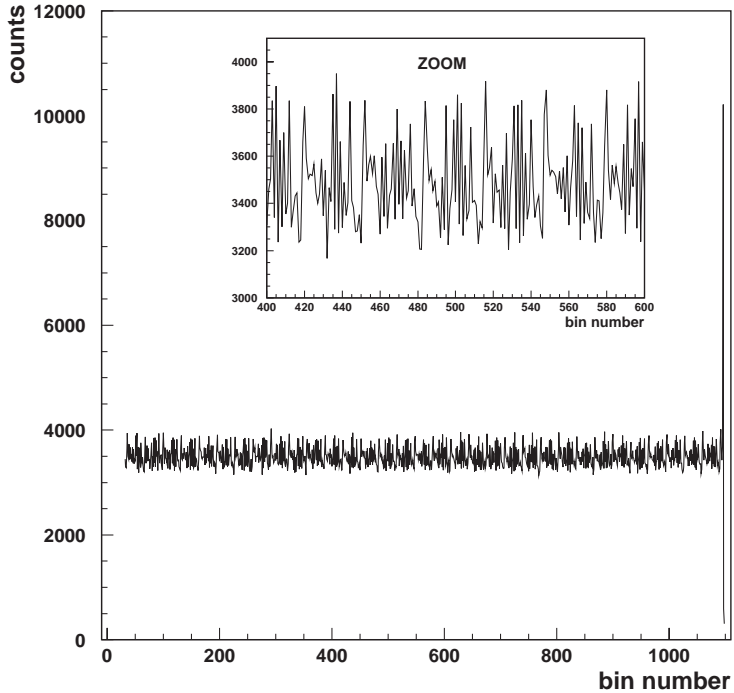, width=.7\textwidth} \caption{White noise
  time distribution for TDC calibration. In the enlargement, the non-linearities of the TDC are seen. Given their high noise level, the first (not shown)
  and last few bins were not used for the recording of physics
  data.}\label{fig:noise}
\end{figure}

The TDC linearity was studied in a dark room by recording a white noise spectrum with high
statistics; in the absence of LED flashes, the
PMT stop signals were provided by the background created by a
controlled light leak. A typical white noise spectrum is displayed in
figure~\ref{fig:noise}, showing the non-linearities associated with
the 32 channel subdivision pattern. A negative slope in the data is
expected, due to the fact that the TDC is single-hit (i.e. stopped by
the first PMT signal) and thus cannot record the arrival time of
further photons; earlier hits are favoured over later ones.
The probability $p_i$
that a hit is recorded in bin $i$ given a flat distribution of arrival time is (to first order) 
$p_i = R\;\delta t \times
(1-R\;\delta t)^{(i-1)} \simeq R\;\delta t \times [1-(i-1) \;R \;\delta t]$
where $R$ is the rate of background noise.

\subsection{Experimental procedure}\label{sec:procedure}

The cable lengths are set on board for the desired source-detector
distance (ranging from 15 to 44~m). Two cable lengths are typically used for each choice of
LED voltage.

After the line has reached the sea bed, the acoustic link is
established from the surface ship. The light intensity is adjusted via
the voltage on the LED pulsers to give a detection efficiency of about
1 detected photon per 100 triggers for the shortest source-detector
distance. This ensures that the PMT is working close to the
single photoelectron regime. The detection efficiency can artificially increase because
of intermittent luminescence bursts. The same intensity is then used
for the longer distance. The discriminator threshold is set at a pulse height
value of 0.3 times the amplitude of the single-photoelectron peak, in
the valley between the peak and the noise.  For a given
source-detector distance, several data sets corresponding typically
to 5$\times$10$^6$ triggers each were collected for each of the two
light colours. The overall time needed to perform the
measurements at one source-detector distance including the drop and
recovery of the line is approximately 4 hours.

The various measurements of the water optical properties at the
{\sc Antares} site from 1997 to 2000 are summarised in table~\ref{tab:data},
in chronological order.

\begin{table}[h]
\begin{center}
\begin{tabular}{|c|c|c|l|}
\hline
\multirow{2}{1.2cm} {\centering Source} & $d_{\rm SD}$ & {Time of} & 
\multirow{2}{4cm} {\centering Comment}\\
 & {(in m)} & { year} & \\
\hline \hline
Blue & 6 -- 27 & Dec. 1997 & Different setup (cf. text)\\ \hline
Blue & 24, 44 & July 1998 & 2 source intensities \\
Blue & 24, 44 & March 1999 & standard\\
UV & 15, 24 & July 1999 & standard\\
UV & 24, 44 & Sept. 1999 & standard\\
Blue, UV & 24, 44 & June 2000 & standard\\
Blue, UV & 24 & June 2000 & 400 m above the sea bed \\
\hline
\end{tabular} \caption{Data recorded for the study of the water light 
transmission properties. The standard configuration is: measurement of the arrival time
distribution of photons from a pulsed isotropic source, one source
intensity (same for the two source-detector distances $d_{\rm SD}$),
source sphere located 100~m above the sea bed
(figure~\ref{fig:line}).}
\label{tab:data}
\end{center} \end{table}

The July 1998 data taken with two source intensities I1 and I2 are used
to study the systematics coming from the shape of the source time
distribution, which exhibits a slight dependence on the LED pulser
voltage (16.5~V in the first case, 18~V in the second). The data recorded 
400~m above the sea bed (June 2000) were compared
to the other data recorded at 100~m above sea bed to 
check the water transparency dependence over this depth range, corresponding to the instrumented range of the {\sc Antares} detector.

\begin{figure}[p] \centering
  \epsfig{file=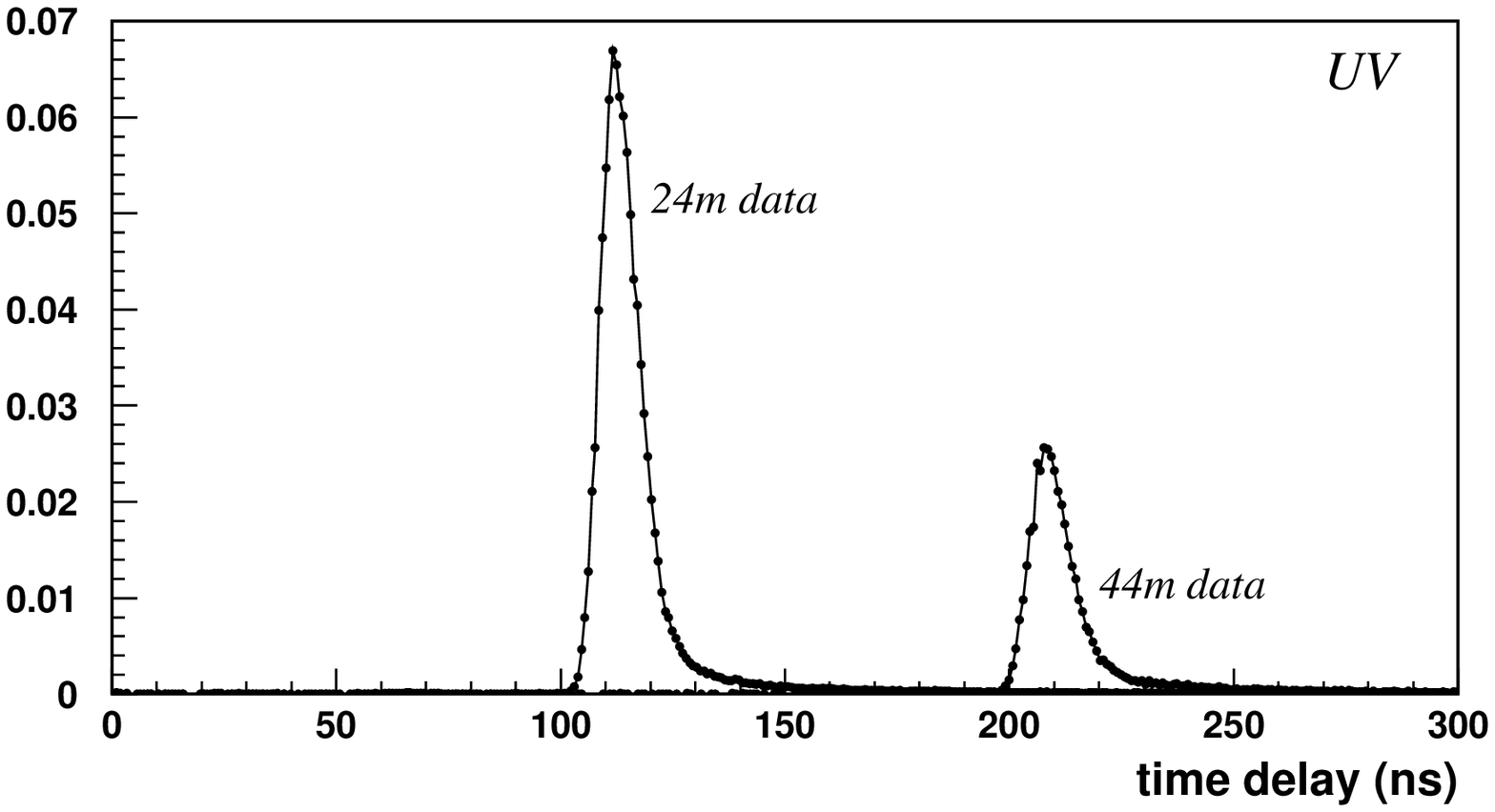, width=.8\textwidth} \caption{Time
  distributions in UV light for the two source-detector distances (24
  and 44~m) of the first June 2000 immersion. Y-axis is proportional to the
  number of photons collected. The 24~m distribution is normalised to unity. The 44~m distribution is normalised with respect to the one at 24~m, in addition to a $(44/24)^2$ factor, so the difference between the two peaks is entirely due to the exponential attenuation factor.}  \label{fig:datalin}
  \epsfig{file=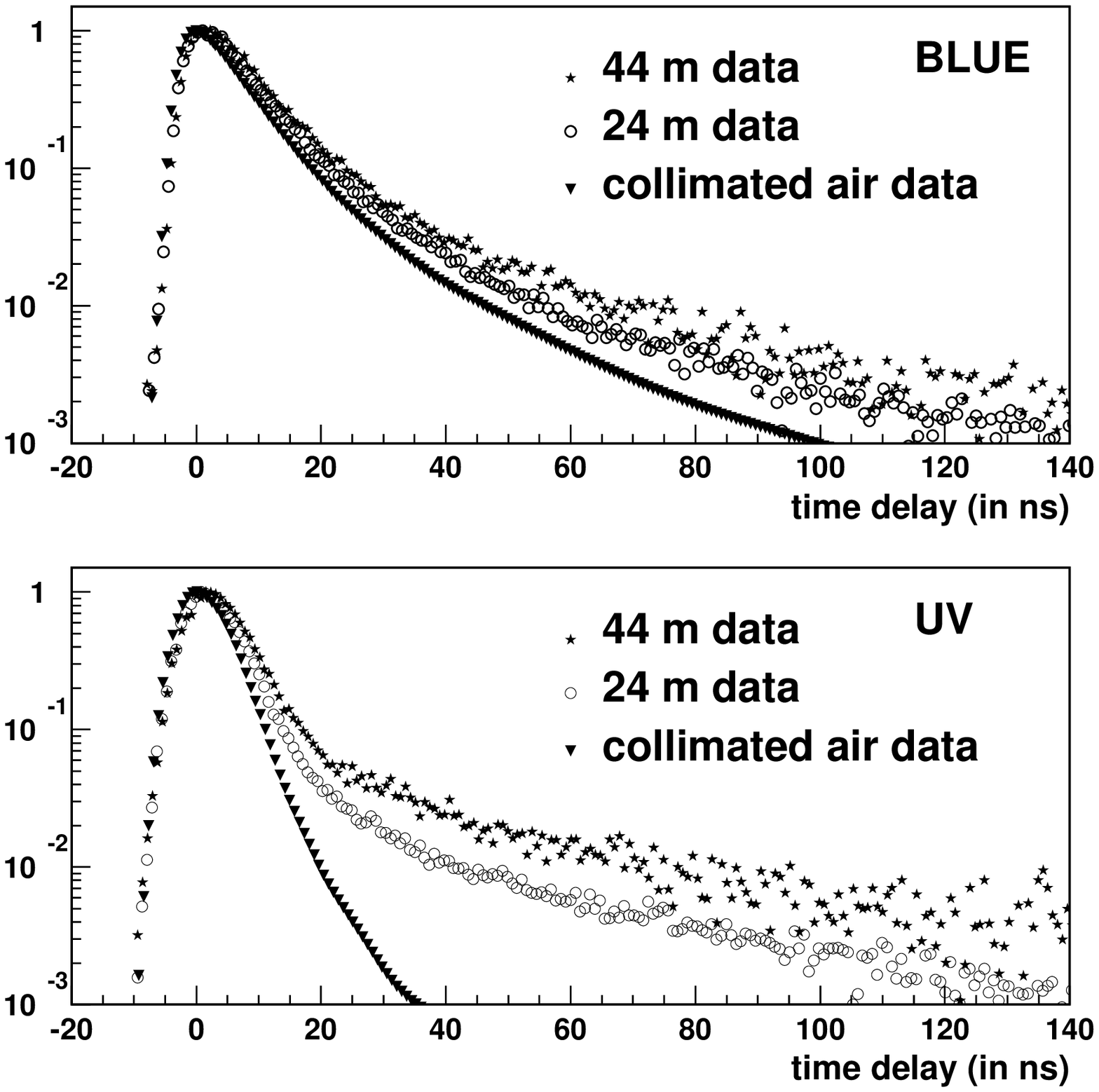,width=.7\textwidth} \caption{Photon arrival times 
  for a collimated air time distribution and for
  two distributions taken {\em in situ}, with source-detector
  distances of 24~m and 44~m.  All distributions are normalised to 
  unity at the peak.  
  Top panel: June 2000 blue data.
  Bottom panel: June 2000 UV data. The origin of the time delay axis is
  set to 0 for direct photons.}  \label{fig:evol}
\end{figure}
As illustrated in figure~\ref{fig:datalin}, all the time
distributions recorded (1998 to 2000) exhibit a small tail of delayed
photons. This corresponds to a small contribution due to scattering between the source and the detector. Figure~\ref{fig:evol}
illustrates, for June 2000 standard blue and UV data, the shape of the photon
arrival time distribution in air (no scattering), and in water with a
source-detector distance $d_{\rm SD}=24$~m or $d_{\rm SD}=44$~m.  Clearly, the width of the main peak comes mostly from
time resolution of the setup. The FWHM of spectra recorded for a
source-detector distance of 24~m is about 10~ns, to be compared with the
intrinsic FWHM of 9~ns of the light source.  As expected, the
scattering tail increases with a larger separation between the source
and the detector.  Scattering is also seen to be more significant in
UV than in blue: slightly larger increase of the width of the peak
region and higher scattering tail, in particular at a distance of
44~m.

The setup deployed for the first immersion (December 1997) was different
from the one described above. An $8''$ photomultiplier tube is located
at a variable distance (6 to 27~m) from a collimated continuous blue
LED source ($\lambda = 466$~nm). Only the integrated intensity was
recorded. The analysis of the data from this immersion is described in
section~\ref{sec:atteff}. All subsequent immersions were done with the
setup described in sections~\ref{sec:line} to \ref{sec:service}.

\section{Simulation of the experiment}\label{sec:sim}

The photon propagation (hence the time distribution of photons at a
distance $R$ from the source) is governed by the following inherent
optical parameters: group velocity of light in the medium $v_g$,
absorption length $\lambda_{\rm abs}$, and volume scattering function
$\beta(\theta) = \widetilde\beta (\theta) / \lambda_{\rm sct}$ with
units $\rm m^{-1}\cdot sr^{-1}$ (where $\widetilde\beta (\theta)$ is
the normalised scattering angle distribution and $\lambda_{\rm sct}$
the scattering length). The scattering function is roughly described
by the scattering length $\lambda_{\rm sct}$ and the average cosine of
the scattering angle distribution (or asymmetry parameter) $\langle \cos
\theta\rangle = 2\pi \int \widetilde\beta(\theta) \,\cos\theta\,
d(\cos\theta)$, under the assumption of a specific shape of the scattering angle distribution.


\subsection{Physics of propagation of light in sea water}\label{sec:phys}

For an isotropic source of photons with intensity $I_0$, the intensity
$I$ detected at a distance $R$ from the source by a PMT with an active
area $A$ is
\begin{equation}
I=I_0\frac{A}{4\pi R^2}e^{-R/\lambda_{\rm att}^{\rm eff}} \; ,
\label{eq:def_latteff}
\end{equation}
where $\lambda_{\rm att}^{\rm eff}$ is the effective attenuation
length, extracted from the total number of photons (i.e. from the integrated
time distributions) recorded for two source-detector distances.

An approximate degeneracy
reduces the number of parameters needed to characterise the time
distribution of photons at a distance $R$ from the source. In
particular, strong correlations can be expected if trying to extract
$\langle\cos \theta\rangle$ and $\lambda_{\rm sct}$ separately, while
$\lambda_{\rm sct}^{\rm eff}$, defined as  
\begin{equation} 
\lambda_{\rm sct}^{\rm eff} \equiv
\frac{\lambda_{\rm sct}}{1-\langle\cos \theta\rangle}  \; ,
\label{eq:defLsctEff}
\end{equation}
describes the main part of the scattering.
\footnote{As a general property of multiple scattering~\cite{Kir}, the
average cosine of the light field produced by a thin narrow parallel
beam after $n$ scattering events $\langle\cos \theta\rangle_n$ is
related to the average cosine for single scattering $\langle\cos
\theta\rangle$ by the relation
\(
\langle\cos \theta\rangle_n=\langle\cos \theta\rangle^n \; .
\)
The average number of scattering events undergone by a photon reaching
a distance $R$ from the source is $n = {L(R)}/{\lambda_{\rm sct}}$
where $L(R)$ is the average path length of these photons. If scattering
is dominantly at small angle, as in natural waters, we have $n\simeq {R}/{\lambda_{\rm
sct}}$. Therefore the average cosine of the light field at distance
$R$ from the source is:
$
\langle\cos \theta\rangle_R \simeq 
\langle\cos \theta\rangle^{R/\lambda_{\rm sct}}.
$
All combinations of $\lambda_{\rm sct}$ and $\langle\cos
\theta\rangle$ that give the same effective scattering length
\[
\lambda_{\rm sct}^{\rm eff} = 
\frac{\lambda_{\rm sct}}{-\ln \langle\cos \theta\rangle}
\]
yield the same $\langle\cos \theta\rangle_R$.  In the case
where $\langle\cos \theta\rangle\simeq 1$, the above relation becomes equation~\ref{eq:defLsctEff}.
}

We describe the scattering angle distribution following the
approach of Morel and Loisel~\cite{Mor}: the scattering angle
distribution is expressed as the weighted sum of molecular and particulate scattering.
The molecular scattering is described by 
the Einstein-Smoluchowski formula for pure water,
\begin{equation}
\widetilde {\beta^m} (\cos\theta) = 0.06225 \; (1+0.835\cos^2\theta)\; ,
\end{equation}
which is reminiscent of the form
\begin{equation}
\widetilde {\beta^{\rm Ray}} (\cos\theta) = 
\frac{3}{16\pi}\;(1+\cos^2\theta) \; ,
\end{equation}
commonly called Rayleigh scattering. The 0.835 factor
(rather than 1) is attributable to the anisotropy of the water
molecules.  
The particulate scattering is described by the Mobley et al.~\cite{Mob} tabulated distribution $\widetilde {\beta^p} (\cos\theta)$, obtained by averaging the similar
particulate angle distributions measured by Petzold in very
different seas~\cite{Pet}, at a wavelength of $514$~nm.

The total normalised scattering angle distribution is of the form:
\begin{equation}
\widetilde\beta(\cos\theta)=\eta\widetilde{\beta^m}(\cos\theta)+
(1-\eta)\widetilde{\beta^p}(\cos\theta) \; ,
\label{eq:sctphasefct}
\end{equation}
with $\eta$ the ratio of molecular to total scattering. The
average cosine of the total scattering angular distribution is
\begin{equation}
\langle \cos\theta \rangle = (1-\eta) \times \langle \cos\theta\rangle^p
= (1-\eta) \times 0.924 \; ,
\label{eq:cos_eta}
\end{equation}
since the average cosine of Petzold's distribution is 0.924.
In natural waters, $\eta$ is typically less than
0.2~\cite{Mor}, so $\langle \cos\theta \rangle$ is large and the
use of equation~\ref{eq:defLsctEff} is justified.


\subsection{The Monte Carlo simulation}\label{sec:MC}

A detailed Monte Carlo simulation, which includes the geometry of the
experimental setup and the optical properties of the medium, has been
used to analyze the experimental time distributions and extract the
light transmission parameters at the {\sc Antares} site. The
parameters of the Monte Carlo are the absorption length $\lambda_{\rm
abs}$, the scattering length $\lambda_{\rm sct}$, the fraction
$\eta$ of molecular scattering, 
the source-detector distances $d_i$, the
origin of time for each distribution (or the time $t_i$ at which
direct photons reach the detector located at a distance $d_i$ from the
source) and the collection efficiency for each distribution.

For each photon, the distance $x$ it will travel before being absorbed
is selected randomly from the probability distribution proportional to
$\exp(-x/\lambda_{\rm abs})$. The photon's distance to its first
scattering is similarly selected according to $\exp(-x/\lambda_{\rm
sct})$. If the absorption distance is shorter, the photon is
propagated to its point of absorption and stopped.  Otherwise, the
type of scattering (molecular or particulate) is selected according to
their respective probabilities $\eta$ and $1-\eta$; the photon is
propagated to its point of scattering where a new photon direction is
sampled from the appropriate angular distribution and a new scattering
distance is drawn. This is repeated until the total length of the
photon path reaches its absorption distance. Time distribution
histograms are filled whenever the photon reaches a radial distance
from the source corresponding to one of the possible source-detector
separations. Weights are applied to take into account the dependence
of the PMT detection efficiency on the angle of incidence of the
photon on the photocathode~\cite{Moo} or to study a possible
anisotropy of the source emission. Each Monte Carlo distribution
results from the propagation of one million photons.

The time distribution of the emitted light pulse is taken from the
one measured in air for direct photons (see figure~\ref{fig:air}), its
angular distribution is taken as isotropic, and its spectrum as
monochromatic (see actual spectral width in figure~\ref{fig:sources})
at its central wavelength.

The obtained Monte Carlo photon spectrum corresponds to a light source
with a vanishingly small intensity, dominated by single photon events. 
For realistic intensity
conditions, since the TDC is working in the single hit mode, one needs
to correct the spectrum for multi-photon events where the first one
only is detected. This correction depends on the average rate of
detection per light pulse trigger, and is calculated assuming Poisson
statistics. The background is removed from the data spectra (see
section~\ref{sec:dataproc} on the data processing) so Monte Carlo spectra are generated with no noise.


\section{Data analysis}


\subsection{Data processing}\label{sec:dataproc}

For each immersion, several data sets were taken with the same
configuration. They are all fully compatible and proved the excellent
reproducibility of the data over a period of a few hours. They are therefore
combined to reduce the statistical noise on the data.

Each recorded time distribution histogram is first corrected for the
non-linearity of the TDC with a bin by bin division by a
slope-corrected white noise time
distribution (cf. section~\ref{sec:service}) as shown in figure~\ref{fig:noise}.

The optical background at this site was studied in detail~\cite{biolum}. 
It consists of a variable bioluminescence
component superimposed on a constant component due to the radioactive decay
of $^{40}$K.  During periods without bioluminescence bursts, it
exhibits a rate $\widetilde{R}$ of about 0.1~kHz/cm$^2$, contributing
as a constant component in the time distribution through random stop
signals. Bioluminescence bursts can reach rates $\widetilde{R}$ up to
several tens of kHz/cm$^2$ generally lasting for hundreds of micro-seconds
to seconds i.e. for longer than the time range of the TDC. Given the highest rates observed, bioluminescence bursts only contribute as an
additional noise which appears as a linearly decreasing component. Extrapolating the result of
section~\ref{sec:service}, the total number of background events $N_i$
in bin $i$ of the spectrum in the region free of LED events is
therefore given by
\begin{eqnarray}
N_i &=& \sum_{j=1}^{N_{\rm triggers}} R_j\;\delta t \;[1-(i-1)R_j\;\delta t] \nonumber \\
 &=& \left(\sum_{j=1}^{N_{\rm triggers}} R_j\;\delta t \right) 
     -(i-1)\left(\sum_{j=1}^{N_{\rm triggers}} (R_j\;\delta t)^2 \right) \nonumber \\
 &=& a-(i-1)b\;, \label{eq:ab}
\end{eqnarray}
where $N_{\rm triggers}$ is the number of triggers and $R_j$ is the
background rate (from bioluminescence and the decay of $^{40}$K)
during cycle $j$ on the $1''$ PMT ($R=\widetilde{R}S$).  The
background contribution is determined by a first-order polynomial fit
to the data in the region free of LED hits (i.e. before the signal
from the LED direct photons): $N_i=a-(i-1)b$. Over 500 bins are
available for the noise fit. The background rates were seen to vary between 0.2 and 2.5~kHz/cm$^2$.


When taking into account the fact that the multi-hit correction
applies to the sum of the hits from the background noise and from the
LED source, the background to be subtracted from each bin $i$ of the data
is given by the following equation (which results from the difference between the expected distributions in the presence and in the absence of noise):
\begin{eqnarray}
N_{\rm noise} (i) &=& 
\sum_{j=1}^{N_{\rm triggers}} \left[(R_j\;\delta t + R_{{\rm LED}_i}\;\delta t)
 \left(1-(i-1)R_j\;\delta t - \sum_{k=1}^{i-1}R_{{\rm LED}_k}\;\delta t \right)\right]
  \nonumber \\
& & - N_{\rm triggers}\times R_{{\rm LED}_i}\;\delta t \left(1-\sum_{k=1}^{i-1}R_{{\rm LED}_k}\;\delta t \right) \nonumber \\
&=& a-(i-1) b - a\sum_{k=1}^{i-1}R_{{\rm LED}_k}\;\delta t -(i-1)\;a\;R_{{\rm LED}_i}\;\delta t
\label{eq:noise}
\end{eqnarray}
where $a$ and $b$ are the same as in equation~\ref{eq:ab} and $R_{{\rm LED}_i}$ is the rate of hits in bin
$i$ coming from the LED source (independent of the cycle
number). Three correcting terms to the canonical value $a$ of the
background appear in equation~\ref{eq:noise}. The slope $b$ in the
noise remained small ($b<10^{-3}$). The intensity of the
source was chosen so as to minimise the multi-hit correction (and
maintain a reasonable signal-to-noise ratio), so that
$\sum_{k=1}^{i-1}(R_{{\rm LED}_k}\;\delta t)$ is at most $\sim 5\%$
and the second correcting term also remains small. The last term,
however, can be quite large in bins where the rate of hits from the LED
source is large. 

\begin{figure}[h] \centering
  \epsfig{file=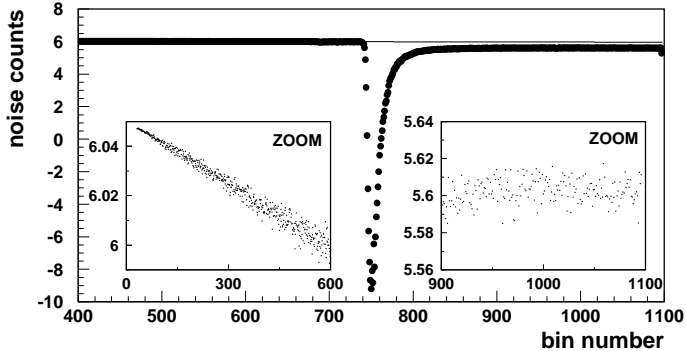, width=.7\textwidth} \caption{Noise counts subtracted from the blue data of June 2000. The thin solid line is the level the noise would have had in the absence of LED hits.}
  \label{fig:noisecontrib} 
\end{figure}

The result of this procedure is illustrated in figure~\ref{fig:noisecontrib} with the example of blue data taken in June 2000. 
The spectrum results from a total of $6\times 10^6$ triggers, with an average collection efficiency of 6\%. The noise level estimated from the first 700 bins is $a=6.05$ with a slope $b=-8.3\times 10^{-5}$. Bin number 750 contains the highest signal, with 3335 counts and a noise contribution of $-9$ hits, meaning that the photons collected earlier in the spectrum prevented 9 signal counts from reaching this bin. The noise level in the tail of the spectrum is stable at a level of 5.6~counts.

After the subtraction of the background, when two different
source-detector distances are available for the same date and source
intensity, a measure of the effective attenuation length is obtained
from the integrated time distributions (cf. section~\ref{sec:atteff}). The experimental time
distributions are then fitted simultaneously to Monte Carlo
distributions to extract the absorption and scattering parameters (cf. section~\ref{sec:fit}).

 
\subsection{Group velocity of light}\label{sec:vg}


The group velocity of light can be computed from the equation
\begin{equation}
v_g = \frac{c}{n}\times \left( 1+\frac{dn/n}{d\lambda/\lambda}\right) \; ,
\end{equation}
using various empirical models for the index of refraction $n$
evaluated for the parameters of the {\sc Antares} site (pressure
$p=230$~atm, salinity $S=38.44\,^\circ\!/\!_{\circ\circ}$ and
temperature $T=13.2^\circ\rm C$).

\begin{figure}[h] \centering
  \epsfig{file=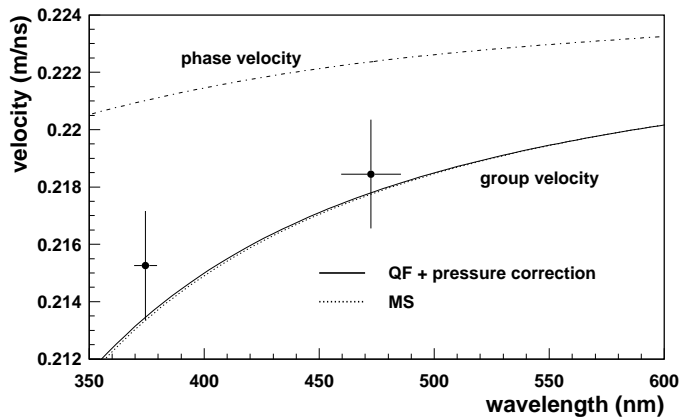, width=.7\textwidth} \caption{Comparison of
  measurements of the group velocity of light with model predictions, for
  $\lambda = 472.5$~nm (Blue) and $\lambda = 374.5$~nm (UV). The phase velocity as a function of wavelength is also shown.}
  \label{fig:index} 
\end{figure}

Using four different experimental data sets of pure water and sea
water under various pressures, Millard and Seaver~\cite{MS}
(thereafter referred to as MS) have developed a 27-term algorithm that
gives the index of refraction to part-per-million accuracy over most
of the oceanographic parameter range (salinity $S=0-40 \,
^\circ\!/\!_{\circ\circ}$, temperature $T=0-30^\circ\rm C$ and
pressure $p=1-1080$~atm), but only over a limited range of wavelengths
($500-700$~nm) so that we need to extrapolate to use it for our
wavelengths of 375 and 473~nm. The result is illustrated in figure~\ref{fig:index}, curve labelled MS.

A simple empirical equation for the index of refraction of sea water
$n(\lambda,S,T)$ for $\lambda$ in $[400-700]$~nm can also be found
in~\cite{Fry} by Quan and Fry (thereafter referred to as QF), based on
data from Austin and Halikas~\cite{Austin}. The pressure dependence
was not included in their equation, so we added it assuming the
same linear dependence as that observed on pressure-temperature plots
from~\cite{livre}. The wavelength dependence of this model is the curve labelled QF in figure~\ref{fig:index}, showing excellent compatibility with the MS model.


As an experimental verification, the following consistency check was performed.
From the time distributions recorded with two source-detector
distances for a unique source intensity, one can extract the group
velocity of light $v_g = \Delta d / \Delta t$ where $\Delta d$ is the
difference between the source-detector distances  for
the two immersions, and $\Delta t$ is the difference between the times
at which the non-scattered photons emitted by the
source reach the detectors located at the two distances from the source.

A length difference $\Delta d = 20.60 \pm 0.14$~m was measured on shore
under a tension of 50~kg on the cables to simulate the buoyancy pull
of the immersed line.\footnote{Parafil cables were seen to stretch by
about 1\% under a tension of 50~kg, although specifications mentioned
a stretch of at most $1.6\,^\circ\!/\!_{\circ\circ}$. Part of the
uncertainty on $\Delta d$ comes from the uncertainty on the
stretchability.} The time differences are
\begin{equation}
\Delta t = \left\{
\begin{array}{lc}
94.3 \pm 0.1 \;({\rm stat.})\pm 0.1 \;({\rm syst.})\;{\rm ns}&({\rm Blue}) \\
95.7 \pm 0.1 \;({\rm stat.})\pm 0.1 \;({\rm syst.})\;{\rm ns}&({\rm UV}) 
\end{array} \right. \; ,
\end{equation}
where the first error is the statistical error and the
second a systematic error coming from the measurement of the
electrical length of the electric cables joining the source to the
detector (a cable was associated with each source-detector distance).
These data therefore imply the following velocities of light (the error
includes the uncertainties on $\Delta d$ and $\Delta t$ stated above), also plotted in figure~\ref{fig:index}:
\begin{equation}
v_g\;({\rm experimental}) = \left\{
\begin{array}{lc}
0.2185 \pm 0.0015\;{\rm m/ns}&({\rm Blue}) \\
0.2153 \pm 0.0015\;{\rm m/ns}&({\rm UV}) 
\end{array} \right.   \; . 
\end{equation}

As can be seen from the figure, experimental and analytical values are in good
agreement. Given the large uncertainty on the determination from the data of the
group velocity of light, however, the value of $v_g$ in the Monte
Carlo is set to the average of the analytical estimates described above, i.e.:
\begin{equation}
v_g\;({\rm model}) = \left\{
\begin{array}{lc}
0.2178\;{\rm m/ns}&({\rm Blue}) \\
0.2134\;{\rm m/ns}&({\rm UV}) 
\end{array} \right.  \; . \label{eq:vg}
\end{equation}
It should be noted that the results of the Monte Carlo fit are not influenced by a small change in the chosen value of the group velocity.

 
\subsection{Effective attenuation length}\label{sec:atteff}

The effective attenuation length $\lambda_{\rm att}^{\rm eff}$ gives
an indication of the fraction of the photons emitted by the source
that are detected (including those that reach the detector although
they have scattered on their way). It varies with the angular
distribution of the source emission and with the angular acceptance of
the detector. It can be computed from the ratio of the total light
detected at the two source-detector distances $d_1$ and $d_2$:
\begin{equation}
\frac{\int N_{d_1}(t)\; dt}{\int N_{d_2}(t)\; dt} = \frac{d_2^2}{d_1^2}\times 
\exp\left(-\frac{d_1-d_2}{\lambda_{\rm att}^{\rm eff}}\right)
\label{eq:l_atteff}
\end{equation}
where $N_{d_i}(t)$ is the time distribution at distance $d_i$ after
background subtraction and multi-photon event correction. The
effective attenuation lengths and the corresponding statistical errors for the data listed in
table~\ref{tab:data} are given in tables~\ref{tab:resultsB} and 
~\ref{tab:resultsUV}.

An uncertainty in the noise subtraction procedure would not affect this result, as explained in section~\ref{sec:noiseuncertainty}. The two data sets taken in 
July 1998 with different LED intensities
yield compatible values of the effective
attenuation lengths ($62.6 \pm 1.0$~m and $60.3\pm 0.4$~m), despite a  large correction for multi-photon events
in the second case.

\begin{figure}[h] \centering
  \epsfig{file=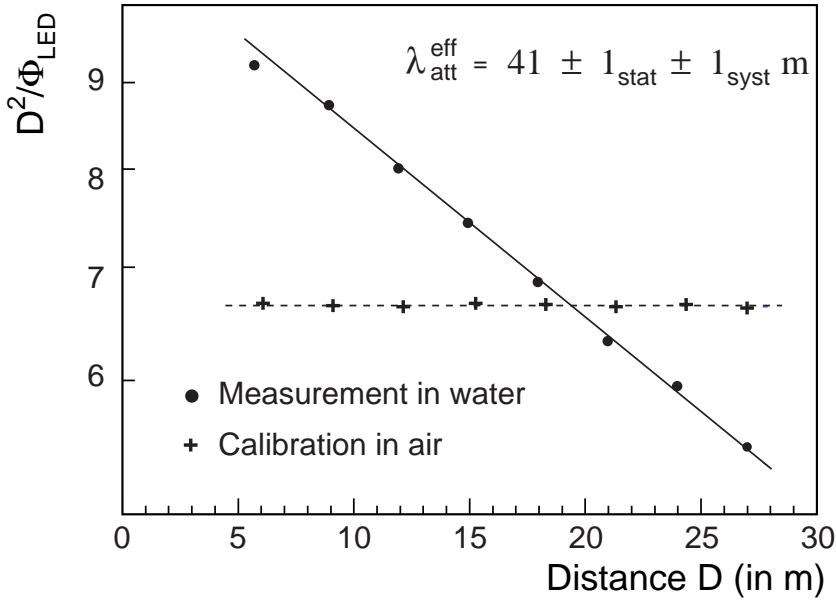, width=.8\textwidth} \caption{Determination of the effective
  attenuation length from the setup immersed in December 1997 (see equation~\ref{eq:test3}).}
  \label{fig:test3} 
\end{figure}

As stated in section~\ref{sec:procedure}, the effective attenuation
length was also measured for the different setup deployed in December
1997, which used a continuous collimated source. While the distance $D$
between the source and the PMT was varied from 6 to 27~m, the
intensity of the source $\Phi_{\rm LED}$ was adjusted so as to yield a
constant current $I_{\rm PMT}$ on the PMT. The setup was calibrated
with a similar procedure in air. The emitted and detected intensities
in water are related by
\begin{equation}
I_{\rm PMT} \propto \frac{\Phi_{\rm LED}}{D^2} \times
{\rm exp}\left(-\frac{D}{\lambda_{\rm att}^{\rm eff}}\right)  \; ,
\label{eq:test3}
\end{equation}
making it possible to estimate the effective attenuation length from
the dependence of the required LED intensity with the distance (cf.
figure~\ref{fig:test3}).  The agreement of the data with a decrease as
given in equation~\ref{eq:test3} yields an effective attenuation
length
\begin{equation}
\lambda_{\rm att}^{\rm eff} \;({\rm Blue,\:collimated})= 41\pm 1\; ({\rm
stat.}) \pm 1\; ({\rm syst.}) \:{\rm m} \; .
\end{equation}
The collimation of the source prevents a direct comparison with values
given in table~\ref{tab:resultsB}.  A Monte Carlo simulation
describing the two setups shows, however, that the above $\lambda_{\rm
att}^{\rm eff}$ would yield $\lambda_{\rm att}^{\rm eff} = 44 \pm 1\;
({\rm stat.}) \pm 1\; ({\rm syst.})$~m with the present (isotropic)
setup, similar to the result found in June 2000.

\subsection{Absorption and scattering lengths}\label{sec:fit}

The uncertainty on the exact cable lengths strongly affects the time
of arrival of direct photons, but has a negligible effect both on the
shape and on the amplitude of the time distributions. The
source-detector distances $d_1$ and $d_2$ are therefore fixed to the
values measured on shore under a tension of 50~kg, while the direct
photons arrival times $t_1$ and $t_2$ are unconstrained to absorb the uncertainties on the distances and avoid biasing
the results. The group velocity of light is taken from
equation~\ref{eq:vg}. The other free parameters of the fit are the
absorption length $\lambda_{\rm abs}$, the fraction of molecular
to total scattering $\eta$ (or equivalently $\langle \cos \theta \rangle$ as
explained in equation~\ref{eq:cos_eta}) and the effective scattering
length $\lambda_{\rm sct}^{\rm eff}$. 

\begin{figure}[h] \centering
  \epsfig{file=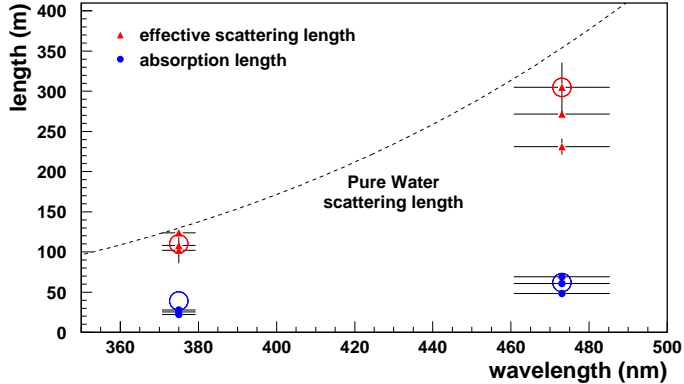,width=.7\textwidth} \caption{Absorption
  (dots) and effective scattering (triangles) lengths
  measured at the {\sc Antares} site at various epochs for UV and blue data.  Horizontal
  error bars illustrate the source spectral resolution ($\pm
  1\sigma$).  The large circles are estimates of the absorption and
  scattering lengths in pure sea water (from~\cite{livre}).  The
  dashed curve is the scattering length for pure water~\cite{Mor74}, upper limit on
  the effective scattering length in sea water.}  \label{fig:result}
\end{figure}

All the results are summarised in tables~\ref{tab:resultsB} and
\ref{tab:resultsUV}, and plotted in figure~\ref{fig:result}. The parameters $\lambda_{\rm abs}$,  $\lambda_{\rm sct}^{\rm eff}$ and $\eta$  result from the global fit described above. The effective attenuation length $\lambda_{\rm att}^{\rm eff}$ is computed according to the method described in the previous section. Therefore, the usual equality between $1/\lambda_{\rm att}^{\rm eff}$ and the sum of the inverses of $\lambda_{\rm abs}$ and $\lambda_{\rm sct}^{\rm eff}$ does not hold, as they were derived from different methods. The scattering length $\lambda_{\rm sct}$ is obtained from $\lambda_{\rm sct}^{\rm eff}$ and $\eta$  according to equations \ref{eq:defLsctEff} and~\ref{eq:cos_eta}. 

The March 99 data were recorded with a low source intensity, and thus a low collection efficiency, resulting in a high statistical error on the measured parameters. The data
sets with two different light intensities available for the July 1998
immersion yield fully consistent results (to within $1\sigma$) for the
fit parameters. Only the mean values are therefore reported here.

\begin{table}[h]
\begin{center}
\begin{tabular}{|c|c|ccc|c|}
\hline {Epoch} & 
\begin{tabular}{c} $\lambda_{\rm att}^{\rm eff}$\\ {(in m)} \end{tabular} 
& 
\begin{tabular}{c} $\lambda_{\rm abs}$\\ {(in m)} \end{tabular} & 
\begin{tabular}{c} $\lambda_{\rm sct}^{\rm eff}$\\ {(in m)} \end{tabular} 
& $\eta$ & 
\begin{tabular}{c} $\lambda_{\rm sct}$\\ {(in m)} \end{tabular} \\ 
 \hline 
July 1998 & $60.6\pm 0.4$ & $68.6\pm 1.3$ & $265\pm 4$ 
& $0.17\pm 0.02$ & $62 \pm 6$\\
March 1999 & $51.9\pm 0.7$ & $61.2\pm 0.7$ & $228 \pm 11 $ 
& $0.19\pm 0.05$ & $58 \pm 18$\\ 
June 2000 & $46.4\pm 1.9$ & $49.3\pm 0.3$ & $301 \pm 3 $ 
& $0.05\pm 0.02$ & $38 \pm 8$\\ \hline
\end{tabular} \caption{Summary of the results for the blue data (statistical 
error only).}
\label{tab:resultsB}
\vspace{.3cm}
\begin{tabular}{|c|c|ccc|c|}
\hline
{Epoch} & 
\begin{tabular}{c} $\lambda_{\rm att}^{\rm eff}$\\ {(in m)} \end{tabular} 
& 
\begin{tabular}{c} $\lambda_{\rm abs}$\\ {(in m)} \end{tabular} 
& 
\begin{tabular}{c} $\lambda_{\rm sct}^{\rm eff}$\\ {(in m)} \end{tabular} 
& $\eta$& 
\begin{tabular}{c} $\lambda_{\rm sct}$\\ {(in m)} \end{tabular} \\
\hline 
July 1999 & $21.9\pm 0.8$ & $23.5\pm 0.1$ & $119 \pm 2 $
&  $0.16\pm 0.03$ & $27 \pm 4$\\ 
Sept. 1999 & $22.8\pm 0.3$& $25.6\pm 0.2$ & $113 \pm  3 $
&  $0.18\pm 0.01$ & $28 \pm 1$\\ 
June 2000 & $26.0\pm 0.5$& $28.9\pm 0.1$ & $133 \pm 3  $
&  $0.12\pm 0.01 $ & $24 \pm 1$\\ 
\hline
\end{tabular} \caption{Summary of the results for the UV data (statistical 
error only). }
\label{tab:resultsUV}
\end{center} \end{table}

\begin{figure}[h] \centering
  \epsfig{file=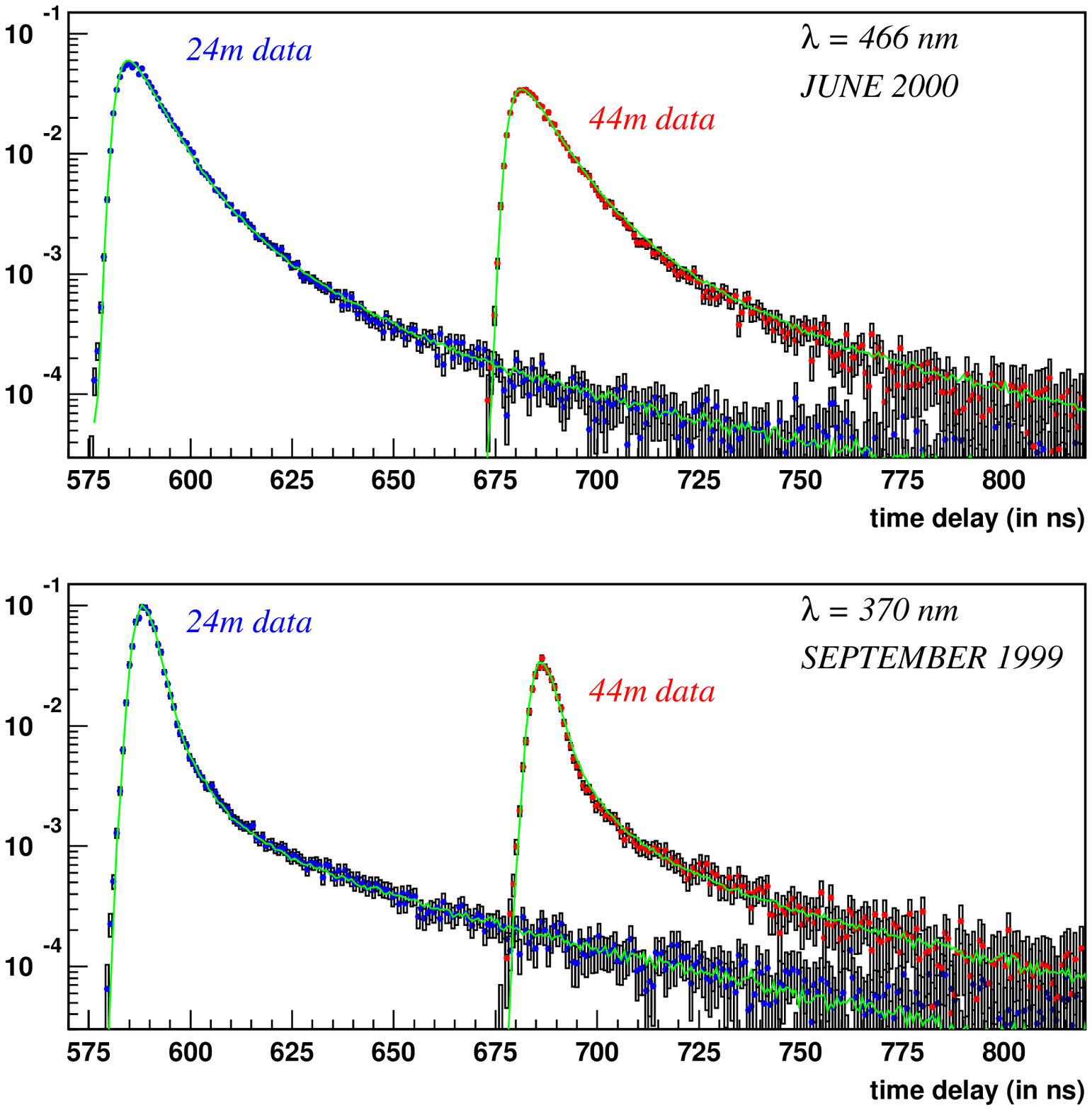,width=.8\textwidth} \caption{Distributions of
  photon arrival times with the best-fit Monte Carlo curves superimposed on
  top of the data. The same normalisation procedure as in figure~\ref{fig:datalin} is applied.
  Top panel: blue data recorded in June 2000. Bottom
  panel: UV data recorded in September 1999.}  \label{fig:mc}
\end{figure}

All fit $\chi^2$'s per degree of freedom are about 1 (within $\pm
0.5$).  Figure~\ref{fig:mc} illustrates, for June 2000 blue and September 1999 UV
data, the photon arrival time distribution in water with a
source-detector distance $d_{\rm SD}=24$~m or $d_{\rm SD}=44$~m, with
the Monte Carlo fit superimposed on top of the two {\em in-situ}
distributions. 

To illustrate the impact of the water transparency properties at various epochs, time distributions have been generated with each set of best-fit parameters assuming a unique setup, and in particular, a unique choice of the source time distribution: that of the June 2000 UV data. The distributions obtained are shown in figure~\ref{fig:scat_art}. 
\begin{figure}[h] \centering
  \mbox{Available as a separate figure : scat\_art.jpg} \caption{Simulated distributions of
  photon arrival times assuming source-detector distances of 24 and 44~m and the experimental setup of June 2000 UV data , for the best fit parameters of each set of data. The same normalisation procedure as in figure~\ref{fig:datalin} is applied.
Top panel: Blue results, bottom panel, UV results.}  \label{fig:scat_art}
\end{figure}

It can be seen in the figure that the absorption length in UV is smaller than in blue (lower relative height, in UV, of the distribution at the largest distance compared to that at the shortest). The higher tail of delayed photons for the UV distribution is compatible with a smaller scattering length in UV. The UV distributions show little dispersion. The first two blue distributions (July 1998 and March 1999) are quite similar. Only the June 2000 blue distribution differs significantly from the other two, with a much lower tail of delayed photons due to its much larger effective scattering length. Its statistical error bars, however, are also very large, due in particular to the long tail of the time distribution of the June 2000 blue source in the calibration spectrum, which reduces the significance of the tail of delayed photons in the data. 

Effective attenuation length results from a combination of absorption
and scattering. Since the scattering tail is small (confirmed by the
large value of $\lambda_{\rm sct}^{\rm eff}$ compared to $\lambda_{\rm
abs}$), the absorption length should be close to the effective
attenuation length. As expected, $\lambda_{\rm att}^{\rm eff}
\stackrel{<}{_\sim} \lambda_{\rm abs}$ in all data sets. The
correlated variations of $\lambda_{\rm att}^{\rm eff}$ and
$\lambda_{\rm abs}$ --- quantities determined independently and quite
robustly --- strengthen the hypothesis of fluctuations of the medium optical properties with time. 

As explained in section~\ref{sec:phys}, $\lambda_{\rm sct}^{\rm eff}$ is used instead
of the more physical $\lambda_{\rm sct}$ in order to avoid large
($\sim 80\%$) correlations between $\lambda_{\rm sct}$ and $\eta$. The
correlation coefficient between $\lambda_{\rm sct}^{\rm eff}$ and
$\eta$ is typically less than $\sim 10\%$.  All other correlation coefficients are compatible with zero, confirming the independence of absorption and scattering properties. 

Despite the comments of section~\ref{sec:phys}, one can try to extract simultaneously the molecular scattering length $\lambda_{\rm sct}^m=\lambda_{\rm sct}/\eta$, 
the particulate scattering length $\lambda_{\rm sct}^p=\lambda_{\rm sct}/(1-\eta)$, and the shape of the particulate phase function using a generic one-parameter Henyey-Greenstein function $\widetilde{\beta^{HG}} (g,\cos\theta)$ :
\begin{equation}
\widetilde{\beta^{HG}}(g,\cos\theta) = 
\frac{1}{4 \pi}\; \frac{1-g^2}{(1+g^2-2g\cos\theta)^{3/2}}  \; ,
\label{eq:beta}
\end{equation} 
where $g = \langle \cos\theta\rangle$. The results are illustrated with the example of the September 1999 UV data. As expected, a fit with these parameters yields very large correlation coefficients between the various scattering parameters :
\begin{equation}
\left[ \begin{array}{crrrr}
&\lambda_{\rm abs} & \lambda_{\rm sct}^m & \lambda_{\rm sct}^p& \langle\cos\rangle^p\\
\lambda_{\rm abs} & 1 &  -0.19 & -0.12 &  -0.08\\
\lambda_{\rm sct}^m & & 1     & -0.87  &  -0.96\\
\lambda_{\rm sct}^p &  &       & 1     & -0.84\\
\langle\cos\rangle^p & &      &        &     1\\
\end{array} \right]\,.
\end{equation}
The plots of figure~\ref{fig:EtaMcEff}  show the existence of a $\chi^2$ minimum as a function of  $\lambda_{\rm sct}^{\rm eff}$, here found for $\lambda_{\rm sct}^{\rm eff} \sim 100$~m, but indicate large  degeneracy between $\eta$ and $\langle\cos\theta \rangle$ (region delimited by the dashed line). For instance, the dot and the star shown in these plots have radically different scattering parameters (cf. table~\ref{tab:res}), although the time distributions for either set of parameters are almost indistinguishable. This follows from  identical scattering angle distributions except for small angles (less than $\sim$30 degrees) to which the experiment is poorly sensitive.
\begin{table}
\begin{center} \begin{tabular}{rcl | rcl}
\multicolumn{3}{c|}{dot position} & \multicolumn{3}{c}{star position}\\
\hline
$\lambda_{\rm sct}$ & = & $30 {\; \rm m}$ &
$\lambda_{\rm sct}$ & = & $76 {\; \rm m}$\\
$\eta$ & = & $0.25 $ &$\eta$ & = & $0.56$\\
$\langle \cos\rangle_{P}$ & = & $0.924$&
$\langle \cos\rangle_{P}$ & = & $0.54$
\end{tabular}\caption{Scattering parameters at two locations of the degeneracy valley indicated with a dot and a star respectively in figure~\ref{fig:EtaMcEff}.} \label{tab:res}
\end{center} 
\end{table}

\begin{figure}[h]
 \centering \mbox{Available as a separate figure : EtaMcEff\_NB.jpg}
  \caption{$\eta$ vs. $\langle\cos\rangle$ for various slices in $\lambda_{\rm sct}^{\rm eff}$. The various shades of grey separate equidistant regions with a difference of 0.5 in the normalised $\chi^2$ per degrees of freedom. The dashed line delimits the $1\sigma$ valley. The dot and the star correspond to the scattering parameters in table~\ref{tab:res}.}   \label{fig:EtaMcEff}
\end{figure}

This degeneracy in the description of the scattering properties explains the use of the widely cited scattering measurements of Petzold~\cite{Pet} for the particulate scattering phase function, only allowing $\eta$, the fraction of molecular to total scattering, to vary.

 
\subsection{Discussion on systematic uncertainties}

Several sources of systematic uncertainties may affect the results of the analysis. These include the slight ani\-so\-tro\-py of the source (see
section~\ref{sec:source}), an uncertainty in the noise determination
(see section~\ref{sec:dataproc}), the reproducibility of the source
intensity at a given voltage for the simultaneous fit of time
distributions recorded several hours apart (for the two
source-detector distances), the knowledge of the time resolution of
the source (time distribution in the absence of scattering, measured
in the lab, cf. section~\ref{sec:source}) and the angular detection efficiency of the detector module. The systematics that affect
the relative normalization of the spectra taken with the two
source-detector distances will mostly have an impact on the effective
attenuation length and on the absorption length, while those that
change the shape of the spectra will rather affect the scattering
length.  These systematics are studied with the help of the Monte
Carlo described in section~\ref{sec:MC}. An estimate of
their effect is given below.

\subsubsection{Source anisotropy}
To test the impact of the 12\% anisotropy of the source, time
distributions are generated with various angular distributions of the
light source emission. While the tail of the time distribution is mostly produced by photons having scattered at large angles, the peak (containing most of the hits) comes from photons that have either reached the detector directly or scattered at small angles, i.e. in either case comes from photons emitted by the portion of the source facing the detector. Since the relevant information for the determination of  $\lambda_{\rm abs}$ and $\lambda_{\rm att}^{\rm eff}$ is the relative normalization of the time spectra collected with two source-detector distances, which to first order is the relative peak height, and since the source sphere remains in the same position with
respect to the detector for all the measurements, the source anisotropy has a negligible impact on the absorption and the effective attenuation lengths. It induces a small distortion on the tail of the time distributions, however, with an impact of the order of 4\% on $\lambda_{\rm sct}$ or $\lambda_{\rm sct}^{\rm eff}$. 

\subsubsection{Noise subtraction }\label{sec:noiseuncertainty}

An uncertainty in the noise subtraction will not affect the absorption
nor the effective attenuation lengths since the noise remains small
compared to the signal, and absorption dominates largely over
scattering. On the other hand, the scattering length is mostly
determined from the tail of the photon arrival time distribution,
which is barely above noise level. The scattering length is therefore
crucially dependent on the precision of the noise subtraction. For
data taken after June 1999, the spectra have 1100 time bins. The noise
contribution before the main signal peak is obtained by a fit over
more than 500~bins with negligible impact of the uncertainty on the
determination of the noise ($<1\%$). The acquisitions of July 1998 and
March 1999, on the other hand, were done with 128 bins only. The noise
is therefore estimated from only $\sim 20$ bins in the worst case
(data with the short source-detector distance). The uncertainty on the
noise determination can then induce an uncertainty on the effective scattering
length of at most 8\%: 16~m,  13~m  and 3~m  for the July 1998,  March 1999  and June 2000 blue results respectively, and  1~m for all the results in UV.

\subsubsection{Stability of LED intensity and PMT efficiency}

The LED intensity and the PMT efficiency are stable over a given immersion, as was verified by
the excellent reproducibility of time spectra taken under the same
conditions up to two hours apart. On two consecutive immersions with
different source-detector distances, however, it is not possible to
check the stability of the setup. This affects the global
normalization of the time spectrum, i.e. the photon collection
efficiency (the larger the intensity, the higher the collection
efficiency), and thus the effective attenuation length (see
section~\ref{sec:atteff}). Depending on the intensity actually used
during each immersion, an assumed uncertainty of 1\% on the LED intensity and PMT efficiency affects
the effective attenuation length (and similarly the absorption length) by 1 to 11\%: 4.5~m, 5.5~m, 1~m and 2~m for the July 1998 (intensity I1), July 1998 (intensity I2), March 1999 and June 2000 blue data respectively, 2~m for the July 1999 and September 1999 UV data and 1~m for the June 2000 UV data. 
The impact on the scattering length is negligible since the shape of the time spectrum is unchanged.

\subsubsection{Source time distribution}

\begin{figure}[h] \centering
  \epsfig{file=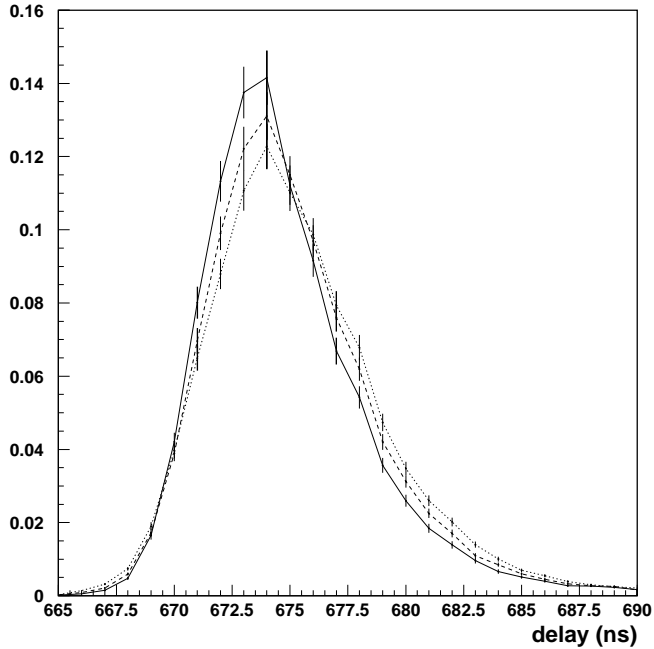,width=.7\textwidth} \caption{Normalised time spectra
  of the UV source for DAC voltages of 4.75V (solid curve), 5V (dashed
  curve) and 5.5V (dotted curve).}  \label{fig:airvsVled}
\end{figure}

The shape of the time distribution of the source varies with the LED input
voltage, as illustrated in
figure~\ref{fig:airvsVled}. Except for the data recorded in June 2000,
time spectra of the source were not always available at the exact same
voltage as the one used for the water measurements. 

Because the relative normalization of spectra at different
source-detector distances is independent of the choice of the
source intensity, the effective attenuation length, and therefore to
first approximation the absorption length, are not affected. The
convolution by a slightly different shape of the source spectrum does
not affect the level of the scattering tail which is set by the
effective scattering length. The latter is therefore also little
affected by a poor knowledge of the source time spectrum. The
strongest effect appears in the peak of the distribution, and 
reflects on the angular dependence of scattering, i.e. on the ratio of
molecular to particulate scattering. More quantitatively, the data
from July 1999 (taken with a DAC voltage of 5.15~V) are fitted with two
different source time spectra, first with one at 4.75~V then with one at 5.5~V. The absorption length is unchanged, the effective scattering length increases by 6~m between the two fits and $\eta$ increases by 0.06. Extrapolating these results to the difference between the voltage used under water and the one used for the measurement of the source spectrum yields an uncertainty of 3~m for the July 1998 and March 1999 blue data and of 1.2~m and 2.4~m for the July 1999  and September 1999 UV data respectively. There is no effect on the June 2000 blue or UV results since the same source voltage is used for calibration and data taking.

\subsubsection{Angular acceptance of the detector sphere}

\begin{figure}[h] \centering
  \epsfig{file=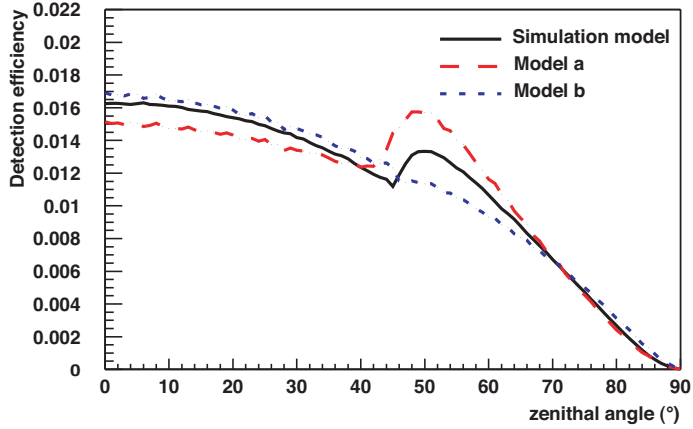,width=.7\textwidth} \caption{Extreme models of the detector sphere angular efficiency (arbitrary units). The solid curve uses a parameterisation of the Moorhead and Tanner data~\cite{Moo} and is the model used in the simulation of the experimental setup. Model a (resp. model b) uses the thinnest (resp. the thickest) of the photocathodes together with the smallest (resp. the largest) complex refractive index.}  \label{fig:deteff}
\end{figure}

The angular detection efficiency (studied in~\cite{Wolf}) of the glass sphere housing the photomultiplier tube was simulated under the assumption of the exact knowledge of various factors, some of which are affected by large uncertainties as for instance the thickness of the photocathode, the thickness of the optical gel which ensures the optical contact between the photocathode and the glass sphere, or the complex refractive index of the photocathode. 
The largest changes in the shape of the efficiency curve are obtained by considering the smallest (respectively largest) value of the complex refractive index within its possible range ($1.10+1.70i$ to $2.75+2.50i$), together with the smallest (resp. largest) value of the photocathode thickness (between 16.4~nm and 26.5~nm). Both of these configurations are shown in figure~\ref{fig:deteff}.

From figure~\ref{fig:deteff} it can be seen that a change in the parameters mentioned above reflects upon the ratio $n_{<45^\circ}/n_{>45^\circ}$ of photons detected at angles smaller or larger than about 45$^\circ$, where the efficiency shows a local enhancement. With model $a$ on the one hand, this ratio decreases, causing more photons to be detected at large angles and therefore requiring a larger scattering length to reproduce the light transmission data. Model $b$ on the other hand requires a shorter scattering length. In both cases, the effect is of the order of 8\% on the effective scattering length.

\subsubsection{Summary of systematic uncertainties}

The total effect of the systematic uncertainties mentioned above is summarised in
tables \ref{tab:systB} and \ref{tab:systUV}, adding all the systematic errors in quadrature.
\begin{table}[h]
\begin{center}
\begin{tabular}{|c|ccc|}
\hline {Epoch} & 
$\lambda_{\rm att}^{\rm eff}${(in m)}& 
$\lambda_{\rm abs}$ {(in m)} & 
$\lambda_{\rm sct}^{\rm eff}${(in m)}  \\ 
 \hline 
July 1998 & $60.6\pm 0.4 \pm 5$ & $68.6\pm 1.3 \pm 5$ & $265\pm 4 \pm 28$ \\ 
March 1999 & $51.9\pm 0.7\pm 1$ & $61.2\pm 0.7\pm 1$ & $228 \pm 11 \pm 24$ \\
June 2000 & $46.4\pm 1.9\pm 2$ & $49.3\pm 0.3\pm 2$ & $301 \pm 3 \pm 27$ \\
 \hline
\end{tabular} \caption{Summary of the results for the blue data (the first error is the statistical error from table \ref{tab:resultsB}, and the second the systematic error).}
\label{tab:systB}
\begin{tabular}{|c|ccc|}
\hline
{Epoch} & 
$\lambda_{\rm att}^{\rm eff}${(in m)}& 
$\lambda_{\rm abs}$ {(in m)} & 
$\lambda_{\rm sct}^{\rm eff}${(in m)}  \\ 
\hline 
July 1999 & $21.9\pm 0.8\pm 2$ & $23.5\pm 0.1\pm 2$ & $119 \pm 2 \pm 10$\\
Sept. 1999 & $22.8\pm 0.3\pm 2$& $25.6\pm 0.2\pm 2$ & $113 \pm  3 \pm 10$\\
June 2000 & $26.0\pm 0.5\pm 1$& $28.9\pm 0.1\pm 1$ & $133 \pm 3 \pm 12 $\\
\hline
\end{tabular} \caption{Summary of the results for the UV data (the first error is the statistical error from table \ref{tab:resultsUV}, and the second the systematic error).}
\label{tab:systUV}
\end{center} \end{table}

The systematic error is significantly larger than the statistical error. It results in an uncertainty of 5 to 11\% on the light transmission parameters. Given these uncertainties, the variations with time of the values of the parameters (cf. section~\ref{sec:atteff}) are reduced to a $\sim 2\sigma$ effect. 
To cope with possible temporal variations of the light transmission parameters, which in this paper are shown to be small, the ANTARES detector will monitor continuously the optical properties at the ANTARES site.
 
 
\subsection{Stability over the line height}

\begin{figure}[h] \centering
  \mbox{\subfigure[Time distributions]
  {\mbox{Available as a separate figure : comp50\_350.jpg} 
  \label{fig: comp50_350}} \quad \subfigure[Distributions ratios]
  {\epsfig{file=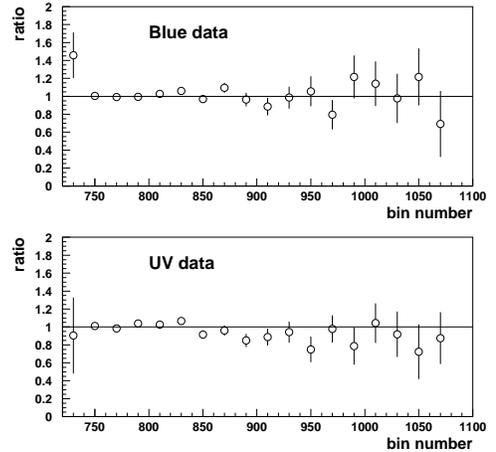, width=.49\textwidth}
  \label{fig: ratio}} }
  \caption{Comparison
  of time distributions recorded 100~m above the sea bed (points with
  grey error boxes in (a)) with the distributions recorded 400~m above the
  sea bed (overlaying black line in (a)). Top panels: blue data, bottom
  panels: UV data. The ratio of the distributions, binned into 16~ns bins, are shown in (b).}  \label{fig:comp50_350}
\end{figure}

During the immersion of June 2000, data were recorded with distances
of 100~m and 400~m between the sea bed and the LED source, in both
blue and UV, to test for a possible variation of the results with
depth.  As illustrated in figure~\ref{fig:comp50_350}, the time
distributions are fully compatible with one another, whether in blue
or in UV, suggesting uniform optical properties along the line height.
The $\chi^2$ between the measurements at the two heights is 297 for
330 degrees of freedom for the blue data, and 252 for 324 degrees of
freedom for the UV data.


\section{Impact on performance of {\sc Antares} detector}

The primary goal of the {\sc Antares} detector is to detect high energy
muons produced by neutrinos interacting around the detector.  In this
section, we investigate the extent to which the detection of muons in
{\sc Antares} will be sensitive to the properties of the water (absorption,
scattering and angular distribution) and how the uncertainties in the
measurement of these properties limit the knowledge of the performance
of the detector.  Since the effect of scattering on the time
distribution is small, scattering is not expected to have a large
impact on the reconstruction efficiency, but it might affect the
angular resolution.

 
\subsection{Event generation and detector simulation}

A sample of $10^{11}$ muon neutrino charged current interactions is
generated for neutrinos with an $E^{-1.4}$ spectrum in the energy range
10 GeV $\leq$ $E_\nu$ $\leq$ 3 PeV. The angular distribution of the
neutrinos is isotropic within the up-going hemisphere. The
interaction point is uniformly distributed within a cylinder of
30~km radius centered on the detector and 30~km height enclosing the
detector and extending downwards from it.  A generator based on LEPTO
\cite{LEPTO} is used for the neutrino interactions. PROPMU
\cite{PROPMU} is used for those events starting outside a 200~m
cylinder surrounding the instrumented volume of the detector to
propagate muons to its surface. Within the 200~m cylinder a full
detector simulation is then performed including the effect of using
different scattering models for the photon propagation. 
The Cherenkov light produced by muons and secondary particles is described as a photon field, subsequently converted into a photomultiplier hit probability. A final step simulates the events in the {\sc Antares} detector.
The detector
geometry used is described in
\cite{Pro}. The results presented here however are not expected to be strongly
dependent on the detector configuration and can therefore be applied to the present configuration~\cite{WEB}.

Two parameterisations of the water scattering properties are
used. The first uses a combination of molecular scattering and
tabulated data from ``Petzold''~\cite{Pet} for particle scattering as
described in section \ref{sec:phys}. The angular distribution is
parameterised by a single parameter $\eta$ (cf.
equation~\ref{eq:sctphasefct}). Two such models, P1 and P2, are
generated with different values of $\lambda_{\rm sct}$ chosen in the
range of observed values (see table~\ref{twater}), with P1
illustrating a conservative case. The second parameterisation uses a
linear combination of two Henyey-Greenstein angular distributions
$\widetilde{\beta^{HG}} (g_i,\cos\theta)$ to approximate the total
scattering angle distribution:
\begin{equation}
\widetilde{\beta} (\cos \theta) = 
\alpha\, \widetilde{\beta^{HG}}(g_1,\cos\theta)\, +\, 
(1-\alpha)\, \widetilde{\beta^{HG}}(g_2,\cos\theta)  \; ,
\label{eq:HG}
\end{equation}
where $\beta^{HG}(g,\cos\theta)$ is given by equation~\ref{eq:beta}.
The parameters describing the angular distribution are then $\alpha$
(the relative contribution of the two HG functions), g1 and g2 (the
$\left< \cos \theta \right>$ of the two HG functions). Five such
models, HG1 to HG5, are generated with different values of
$\lambda_{\rm sct}$ and $\left< \cos \theta \right>$.

The simulated models are given in table \ref{twater}. They have been chosen to
include both a set of ``reasonable'' water properties and extreme
cases to probe the maximum effect of each parameter. The numbers
quoted for these models are at a wavelength corresponding to the blue
LED. The wavelength dependence of the scattering lengths is taken
according to the Kopelevich model \cite{livre}.

\begin{table}[h]
\begin{center}
\begin{tabular}{|c|c|c|c|c|c|c|c|} \hline
Model & $\lambda_{\rm sct} ({\rm m})$ & $\langle\cos\theta\rangle$ & 
$\lambda^{\rm eff}_{\rm sct} ({\rm m})$  & $\eta$ & $\alpha$ & 
g1 & g2  \\ \hline
P1 & 40.8 & 0.77 & 175 & 0.17 & & & \\
P2 & 52.0 & 0.77 & 223 & 0.17 & & & \\
HG1 & 52.0 & 0.77 & 223 & & 1.000 & 0.77 & 0.0 \\
HG2 & 22.3 & 0.90 & 223 & & 1.000 & 0.90 & 0.0 \\
HG3 & 4.4  & 0.98 & 223 & & 1.000 & 0.98 & 0.0 \\
HG4 & 40.8 & 0.90 & 396 & & 0.985 & 0.92 & -0.6 \\
HG5 & 52.0 & 0.90 & 505 & & 0.985 & 0.92 & -0.6 \\ \hline
\end{tabular}
\caption{Simulated water models and parameters (for $\lambda =
466$~nm).}
\label{twater}
\end{center}
\end{table}

The scattering angle distributions of all seven models are shown in the upper panel of figure~\ref{fig:articleDistrib2}. The corresponding time distributions 24~m and 44~m away from the source, as would be measured with the dedicated {\sc antares} setup (configuration of June 2000 with the blue source), are shown in the lower panel. The variety of time distributions is much larger than that actually observed in the data (see for comparison figure~\ref{fig:scat_art}).
\begin{figure}[h] \centering
  \mbox{Available as a separate figure : articleDistrib2.jpg} \caption{Top panel: angular
  distributions $\widetilde\beta(\cos\theta)$ for the models P1, P2, HG1 to HG5 described
  in the text, as well as for pure molecular and pure  particulate scattering described in section~\ref{sec:phys} (grey curves). Bottom panel: time distributions for each of the models at source-detector distances of 24~m and 44~m (same normalisation procedure as in figure~\ref{fig:datalin}). From top to bottom (better distinction on the second set of curves due to the larger distance of propagation): P1, then P2, then HG1, HG2 and HG3 almost indistinguishable from one another, then HG4 and finally HG5.}
  \label{fig:articleDistrib2}
\end{figure}

Models P1 and P2 include both molecular and particulate contributions to the total scattering angle distribution, the molecular part being a major contributor to the delayed signal, due to its backscattering component which is as significant as its forward scattering one. The two models only differ by their scattering lengths. The smaller scattering length of model P1 therefore generates the larger tail of delayed photons.

Model HG1
reproduces the same $\lambda_{\rm sct}$ and $\left< \cos \theta
\right>$ as model P2 but with a different shape for the scattering
angular distribution, as illustrated in the upper panel of figure~\ref{fig:articleDistrib2}. Its lesser backscattering component causes the corresponding time distribution to have less delayed photons. In addition, the peak width is increased.
Models HG2 and HG3 have the same $\lambda^{\rm eff}_{\rm sct}$ as model
HG1 but with different values of $\left< \cos \theta \right>$, probing the impact of the angular distribution. With HG1, HG2 and HG3 distributions similarly dominated by forward scattering, it is the effective scattering length that governs the levels of the tail of delayed photons. The corresponding time distributions are thus, as expected, almost indistinguishable for the three models. Models HG4 and HG5 have the same $\left< \cos \theta \right>$ as model HG2 but with different $\lambda^{\rm eff}_{\rm sct}$. The angular
distribution in models HG4 and HG5 is similar to that of model HG2 with an enhanced backscattering component. The latter is not sufficient however to raise significantly the tail of delayed photons and, as can be expected, the increasing effective scattering lengths lowers the levels of the tail of delayed photons. 

The absorption profile is the same in all cases and corresponds to
that in \cite{Price} normalised to 62.5~m at 470~nm.

Model P2 is the one that best reproduces the experimental
results described in the previous sections.

 
\subsection{Event reconstruction and analysis}

The three-dimensional reconstruction involves several stages of hit
selections to remove PMT hits due to $^{40}$K noise and
bioluminescence, as well as several pre-fits based on plane wave fits through local
coincidence and high amplitude hits. The final step is based on a
maximum likelihood fit to the distribution of photon arrival times
with respect to the expected arrival time of Cherenkov light at a
wavelength of 470~nm. The form of the likelihood function is taken
from an independent Monte Carlo simulation for muons produced by
neutrinos with an $E^{-2}$ spectrum above 1~TeV. That Monte Carlo
includes a 55~m effective attenuation length but neglects scattering so there is no initial assumption towards a
preferred scattering model. Since the likelihood function is greatly
dominated by the peak where the direct photons were expected, the
performance of the reconstruction is not strongly affected by the
shape of the likelihood function.\footnote{The event reconstruction and
selection described here is not optimal for any specific
analysis. Rather, it is a general approach with no strong assumptions
to provide an unbiased assessment of the effect of different
scattering models on the performance.}

The performance of the detector is then defined by two figures of merit of
the reconstruction:

\begin{description}
\item[Angular resolution $\Delta \alpha$] defined as the median angle
between the Monte Carlo neutrino and the reconstructed track.
\item[Effective volume] defined as the fraction of generated events
(per bin) which remain after reconstruction and selection, multiplied by the
generation volume.
\end{description}

Both of these quantities are defined after selection cuts which eliminate misreconstructed tracks and ensure the purity of the data sample.

 
\subsection{The effect of different water models}

\begin{figure}[h] 
  \centering \epsfig{file=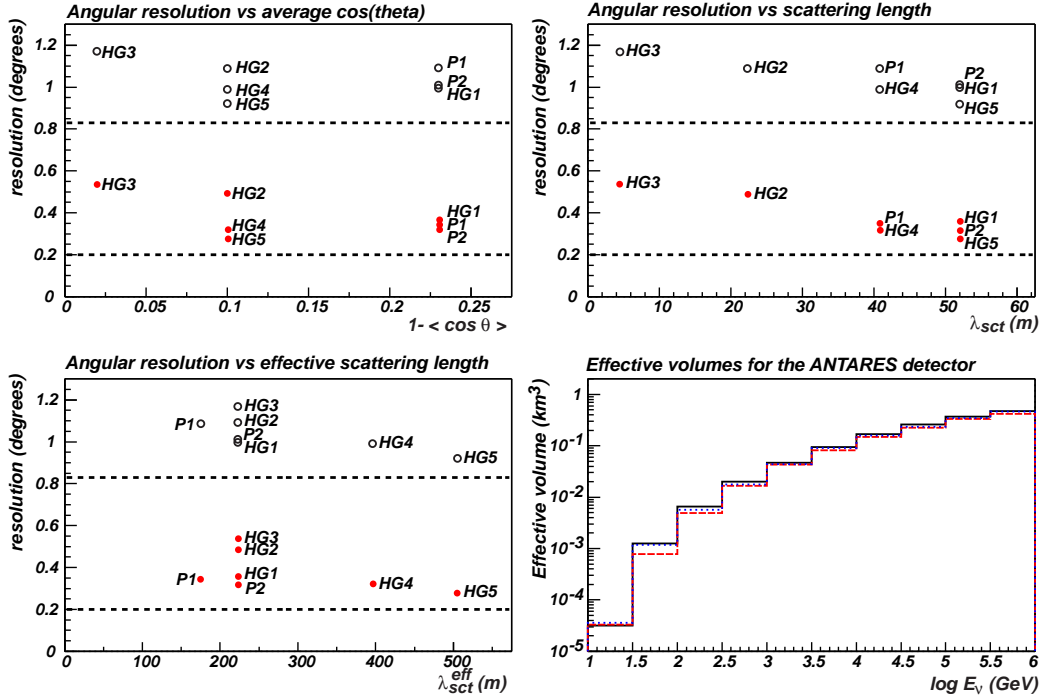,width=\textwidth}
  \caption{Angular resolution for each of the water models as a
  function of $1-\left<\cos\theta\right>$, $\lambda_{\rm sct}$ and
  $\lambda^{\rm eff}_{\rm sct}$. The upper points (open circles)
  correspond to the neutrino resolution for $0.3 < E_\nu < 3$~TeV with
  an $E^{-1.4}$ spectrum, the lower points (filled circles) for $30 <
  E_\nu < 300$ TeV. The horizontal lines show the angular resolution
  obtained in the absence of scattering in each case. The effective
  volumes are shown for the two models with the extreme angular
  resolutions, HG3 (dashed curve) and HG5 (dotted curve) and in the
  absence of scattering (upper solid curve).}  \label{fig:scatperf}
\end{figure}

The angular resolution as defined above is shown for neutrino energies
around 1~TeV ($0.3 < E_\nu < 3$~TeV) and around 100~TeV ($30 <
  E_\nu < 300$ TeV) for each of the simulated water models in figure
\ref{fig:scatperf} as a function of the individual parameters. Several
effects contribute to this resolution.  The angle between the muon and
neutrino at the interaction vertex decreases with increasing neutrino
energy. At 1 TeV, this angle is $0.7^\circ$ on average~\cite{Pro} and
is the most significant contribution to the neutrino angular
resolution, whereas at high energies the muon and neutrino are
essentially collinear so the accuracy of reconstructing the muon track
dominates the angular resolution. The error in the event
reconstruction contributes an additional error, bringing the
resolution at 1~TeV up to $0.8^\circ$. The scattering increases this
further to as much as $1.2^\circ$, depending on the water model. In
the high energy regime, relevant to neutrino astrophysics, the effect
of scattering plays a dominant role. For 100~TeV, the average
muon-neutrino angle is only $0.04^\circ$, the reconstruction brings it
up to $0.20^\circ$ and the scattering further increases it to as much
as $0.53^\circ$. 

The remarkable agreement between the angular resolutions obtained for
models P2 and HG1 implies little sensitivity to the precise shape of
the scattering angular distribution. Same $\left<\cos\theta\right>$ and same $\lambda_{\rm sct}^{\rm eff}$ appear to be sufficient to determine the performance of the detector.

The clearest dependence seen in figure \ref{fig:scatperf} is on the
scattering length $\lambda_{\rm sct}$. A shorter scattering length
degrades the angular resolution. At 1 TeV, the effect of even the most
extreme models considered here is at the level of $\pm 12$\%. It is at
high energies where the differences between the water models in the
neutrino angular resolution become most significant, at the level of
$\pm 30$\% around the central value.

The effective scattering length $\lambda^{\rm eff}_{\rm sct}$ alone is not
enough to describe the effect of scattering on the angular resolution,
as virtually the full range of angular resolutions obtained are seen
for a single effective scattering length of 223~m. Similarly, a wide
range of values is seen for a single value of $\left<\cos\theta\right>
= 0.9$.

Given the results of the {\em in-situ} measurements
(table~\ref{tab:resultsB}), we can reasonably estimate $200 <
\lambda^{\rm eff}_{\rm sct} < 400$~m and a $\left< \cos \theta \right
> \sim 0.75$ for the blue band. With this assumption, the variation in
angular resolution, even at high energy, is at the level of $\pm 10$\%
around the central value of the angular resolution at
$0.32^\circ$. Within this, the effect of assuming different angular
distributions (with the same $\left<\cos\theta\right>$) gives an
uncertainty at the level of $\pm 6$\%. Even for the very conservative
assumption where $\lambda_{\rm sct} > 30$~m, the uncertainty on the
angular resolution only increases to 12\%.

The variation in the effective volume of the {\sc Antares} detector over a
wide range of neutrino energies between the extreme models is at the
level of $\pm 5$\%, with models having the worst angular resolutions
also yielding the lowest effective volume (see figure~\ref{fig:scatperf}).

 
\subsection{Angular resolution of the {\sc Antares} detector}

\begin{figure}[h] \begin{center} 
  \epsfig{file=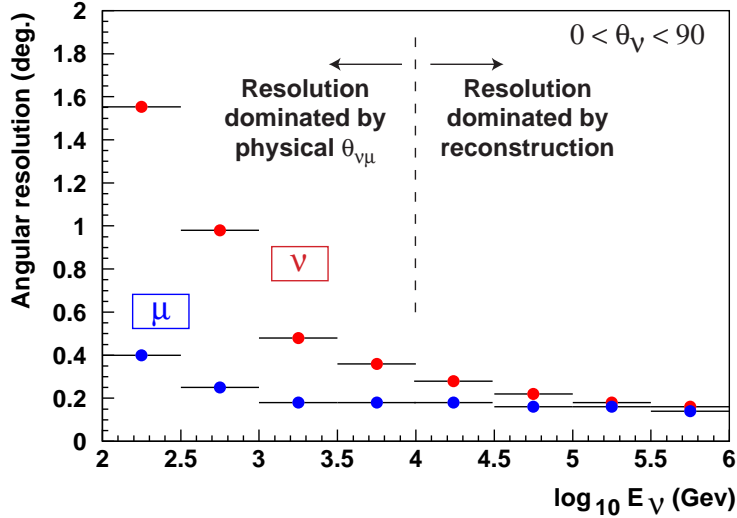,width=.7\textwidth} \caption{Angular
  resolution (for scattering model P2) as a function of the neutrino
  energy for the reconstruction of the muon track (lower curve) or
  that of the parent neutrino (upper curve) for a 10-string
  detector.}  \label{fig:angres} \end{center}
\end{figure}

The present design of the {\sc Antares} detector comprises a 12-string network
that will be immersed over the next few years. The event
selection can be optimised in terms of angular resolution and
effective volume of the detector. The track is obtained by a 2-stage
fit: the position and orientation of the PMTs that have been hit are
used to obtain points that the track is likely to have crossed. These
points are used to obtain an initial track fit. The most probable
track is then obtained by the minimization of a function involving the
residuals of the times at which the Cherenkov photons emitted along
the track reach the PMTs of the detector~\cite{carmona}.
  In the present stage of the
reconstruction software, and considering the
scattering model that most closely reproduces the data presented in
this paper, model P2 described above, the angular resolution for up-going muon
tracks is illustrated in figure~\ref{fig:angres}. For energies $E_\mu > 300$~GeV the angular resolution for a $E^{-1.4}$ spectrum is
\begin{eqnarray}
\Delta \alpha {\rm (\mu)} &=& 0.20^\circ \pm 0.01^\circ \;({\rm stat}) 
                              \pm 0.02^\circ\;({\rm syst}) \\
\Delta \alpha {\rm (\nu)} &=& 0.32^\circ \pm 0.02^\circ \;({\rm stat}) 
                              \pm 0.04^\circ\;({\rm syst}) \:\:.
\end{eqnarray}
The systematics are
computed from the study presented in the previous section.


\section{Conclusions}

The light transmission at the {\sc Antares} site has been studied
intensively with dedicated setups designed by the
collaboration. Absorption and scattering properties of the water for
blue light ($\lambda=473$~nm) and UV light ($\lambda=375$~nm) were
obtained by measuring the distribution of the arrival times of photons
emitted by a pulsed LED source and collected several tens of meters
away by a fast photomultiplier tube.

The group velocity of light is found in good agreement with
predictions from analytical models. The absorption length is seen to
vary slightly in time, with typical values of 60~m in blue and 25~m in
UV. These values allow a large effective area ($\sim 0.1\,{\rm km}^2$
for neutrino-induced muons with energies in the PeV range) for the
planned 12-string {\sc Antares} detector. With the angular distribution of
scattering modelled following the standard approach of oceanographers,
the scattering length $\lambda_{\rm sct}$ can be extracted with good confidence from the
data, yielding an effective scattering length $\lambda_{\rm sct}^{\rm
eff} = \lambda_{\rm sct} / (1-\langle\cos \theta\rangle)$ of $\sim
260$~m in blue and $\sim 120$~m in UV. The various parameters describing the light transmission properties are affected by a 5 to 11\% uncertainty, dominated by systematics. 
Given these large scattering
lengths, an angular resolution of $0.3^\circ$ should be achieved for
$E_\mu > 300~\rm GeV$, according to the present status of the reconstruction
software. The uncertainty in the knowledge of the water properties
(due for instance to the observed variations) affects our knowledge of the
angular resolution and effective volume of the detector by 10\% and 5\%
respectively.

The light transmission properties were checked to be constant at a
given time at the two extreme levels (100~m and 400~m above the sea
floor) of the active part of a detector line.

The water properties will be monitored with {\em in-situ} dedicated
instruments during the lifetime of the {\sc Antares} detector so that the
instantaneous values will be available for use in the muon
reconstruction software.


\end{document}